\documentclass{nature}

\usepackage{amsmath} 

\usepackage{graphicx}
\usepackage[font={small}, singlelinecheck=false]{caption}
\usepackage{array}
\usepackage{chngpage}
\usepackage[euler]{textgreek}

\makeatletter
\let\saved@includegraphics\includegraphics
\AtBeginDocument{\let\includegraphics\saved@includegraphics}
\renewenvironment*{figure}{\@float{figure}}{\end@float}

\DeclareCaptionLabelSeparator{line}{ {\vline height 8pt depth 0.5pt width 1pt} }
\usepackage[labelsep=line]{caption}
\makeatother
\begin{document}

\title{Monopolar and dipolar relaxation in spin ice Ho$_2$Ti$_2$O$_7$}

\author{Yishu Wang$^{1,2,*}$, T. Reeder$^1$, Y. Karaki$^3$, J. Kindervater$^1$, T. Halloran$^1$, N. Maliszewskyj$^2$, Yiming Qiu$^2$, J. A. Rodriguez$^{2,4}$, S. Gladchenko$^2$, S. M. Koohpayeh$^{1,5}$, S. Nakatsuji$^{1,6,7,8}$ \& C. Broholm$^{1,2,5,*}$}

\maketitle

\begin{affiliations}
 \item Institute for Quantum Matter and Department of Physics and Astronomy, Johns Hopkins University, Baltimore, Maryland 21218, USA
 \item NIST Center for Neutron Research, National Institute of Standards and Technology, Gaithersburg, Maryland 20899, U.S.A
 \item Faculty of Education, University of the Ryukyus, Nishihara, Okinawa 903-0213, Japan
 \item Department of Materials Science and Engineering, University of Maryland, College Park, Maryland 20742, USA
 \item Department of Materials Science and Engineering, Johns Hopkins University, Baltimore, Maryland 21218, USA
 \item Institute for Solid State Physics, University of Tokyo, Kashiwa, Chiba 277-8581, Japan
 \item Department of Physics, University of Tokyo, Hongo, Bunkyo-ku, Tokyo 113-0033, Japan
 \item Trans-scale Quantum Science Institute, University of Tokyo, Hongo, Bunkyo-ku, Tokyo 113-0033, Japan
 \item[*] Correspondence: wangyishu@jhu.edu, broholm@jhu.edu
 \item[] \today
\end{affiliations}

\noindent
\textbf{When degenerate states are separated by large energy barriers, the approach to thermal equilibrium can be slow enough that physical properties are defined by the thermalization process rather than the equilibrium. The exploration of thermalization pushes experimental boundaries and provides refreshing insights into atomic scale correlations and processes that impact steady state dynamics and prospects for realizing solid state quantum entanglement. We present a comprehensive study of magnetic relaxation in Ho$_2$Ti$_2$O$_7$ based on frequency-dependent susceptibility measurements and neutron diffraction studies of the real-time atomic-scale response to field quenches. Covering nearly ten decades in time scales, these experiments uncover two distinct relaxation processes that dominate in different temperature regimes. At low temperatures (0.6~K$\mathbf{<T<}$1~K) magnetic relaxation is associated with monopole motion along the applied field direction through the spin-ice vacuum. The increase of the relaxation time upon cooling indicates reduced monopole conductivity driven by decreasing monopole concentration and mobility as in a semiconductor. At higher temperatures (1~K$\mathbf{<T<}$2~K) magnetic relaxation is associated with the reorientation of monopolar bound states as the system approaches the single-spin tunneling regime. Spin fractionalization is thus directly exposed in the relaxation dynamics. }

A rare-earth atom in a crystalline solid can form an isolated Ising-like doublet with large angular momentum $\pm\hbar J$ where low temperature reversal only occurs through transverse-field-driven quantum tunneling~\cite{paulsen1995quantum}. At elevated temperatures the crystal-field energy barrier separating reversed states can be surmounted over a time scale $\tau$ following the Arrhenius form~\cite{paulsen1995quantum, ehlers2006dynamics, ruminy2017phonon}:
\begin{equation}
\tau(T)=\tau_0\exp(\Delta/T).
\label{Arrhenius_law}
\end{equation}
The effective barrier $\Delta$ is near the first excited crystal-field level, while  $\tau_0$ ($\simeq10^{-12}$ s for $\rm Ho_2Ti_2O_7$~\cite{ehlers2006dynamics}) is the lifetime of the crystal-field ground state.

The notoriously slow dynamics of Ising spins is compounded when they form a macroscopic system with competing interactions that define a complex disconnected energy landscape. This is the case in classical spin ice~\cite{ramirez1999zero} where Ising spins occupy the vertices of corner-sharing tetrahedra in the cubic pyrochlore structure (Fig. 1a-c). Net ferromagnetic nearest-neighbor interactions ($J_\text{eff}$) resulting from a combination of superexchange and dipolar interactions define a degenerate manifold of states with Pauling entropy $\Delta S\approx \frac{1}{2}R\ln \frac{3}{2}$~\cite{ramirez1999zero,melko2004monte}. AC susceptibility measurements in $\rm Dy_2Ti_2O_7$ show spin relaxation times increase beyond the single-spin quantum tunneling time scale of $\sim 10^{-5}$~s upon cooling below 2 K and exceed $10^5$~s for $T<0.3$~K where the spin system falls out of thermal equilibrium trapped in a spin ice state with near perfect local order but unable to achieve long range order~\cite{matsuhira2001novel, snyder2004low, matsuhira2011spin,  yaraskavitch2012spin}.

In this letter, we report the discovery of a distinct thermal cross over in the magnetic dynamics of $\rm Ho_2Ti_2O_7$ spin ice from a low temperature regime with a well defined Debye-like relaxation, to a higher temperature regime with a broader and faster relaxation spectrum. Enabled by a new class of ultra-pure $\rm Ho_2Ti_2O_7$ single crystals~\cite{ghasemi2018pyrochlore}, a novel time-resolved magnetic neutron scattering technique, and broad-band AC magnetometry and susceptometry, our experiments resolve  apparent discrepancies in the literature where experiments probing different temperature regimes in more disordered samples probed one or the other but never both of these distinct relaxation regimes~\cite{quilliam2011dynamics,eyvazov2018common}. The totality of the data allows the association of the two regimes with the collective dynamics of monopoles and individual dipoles respectively.

Fig. 1d shows elastic magnetic neutron scattering ${\cal S}(\bf Q)$ in the $(HHL)$ scattering plane deep in the spin-ice regime of $\rm Ho_2Ti_2O_7$ ($T= 0.95$~K). Devoid of Bragg peaks, the exclusively diffuse magnetic scattering surrounding high-symmetry points on the Brillouin zone boundary, such as (001), (003), and $(\frac{3}{2}\frac{3}{2}\frac{3}{2})$ reflect local non-collinear ferromagnetic Ising spin order on each tetrahedron (Fig. 1a). The faint ``bow tie"-like pinch points at the (002) and (111) zone centers indicate the divergence free nature of the magnetization in the spin ice ensemble~\cite{fennell2009magnetic, morris2009dirac}. 

Fig. 1e shows the impact on scattering of applying $H=250$~Oe along $\langle1\bar{1}0\rangle$ and perpendicular to the $(HHL)$ scattering plane. Resolution-limited magnetic Bragg peaks form at (002) and (220), with intensity proportional to the induced magnetization squared. There is also a general reduction in diffuse scattering throughout the $(HHL)$ plane that is  $\bf Q$-independent, uncorrelated with ${\cal S}({\bf Q})$ to within statistical accuracy, and not associated with changes in the intensity of pinch point scattering (see Supplementary Information Note 1 and Fig. S1). These observations negate field-induced changes in the monopole density and imply the net magnetization is induced through motion of magnetic monopoles (Fig. 1b,c) perpendicular to the scattering plane, whence along the field direction. Changes in the $\bf Q$-dependence of ${\cal S}({\bf Q})$ for $\bf Q\perp H$ were previously shown to require considerably larger magnetic fields\cite{clancy2009revisiting}.

Fig. 1f-h show the time-dependence of the coherent and diffuse magnetic scattering. The resolution-limited nature of the magnetic Bragg peak indicates the induced magnetization is homogeneous throughout the coherence volume of the scattering experiment, which extends over $\sim (500~\AA)^2$ perpendicular to the applied field but just 9~\AA\ along the field direction (due to the vertically focusing monochromator and analyzer). The integrated magnetic Bragg intensity at ${\bf Q}=$ (002) (Fig. 1h) is a measure of the time-dependent transverse-to-$\bf Q$ magnetization squared $I_{\text{(002),mag}}(t) \propto M_\perp^2(t)$. For comparison and in the same units is shown the time dependence of the diffuse magnetic scattering throughout the accessible region in Fig. 1e excluding Bragg peaks. As anticipated from the total moment scattering sum-rule, what is gained in magnetic Bragg scattering while the field pulse is on, is lost in diffuse scattering that is independent of $\bf Q\perp H$. 

Having established that time-dependent Bragg intensity at (002) is a measure of $M_\perp^2(t)$ and hence coherent monopole displacement along ${\bf H}\perp (002)$, we can extract the time scale over which monopole drift ceases following a step change in magnetic field by analyzing $I(t)$ data sets such as those shown in Fig. 1h and Fig. 2a. Upon cooling from 1.3 K to 0.6 K, this time scale increases from milliseconds to hours. A stretched exponential description of $M(t)$ (see also Methods) provides a good account of the intensity data (solid lines in Fig. 1h and Fig. 2a) and the corresponding $T-$dependent time constants $\tau$ and exponent $\beta$ are shown with solid black symbols in Fig. 3b and 3c, respectively.    

Sampling the magnetic relaxation in the frequency-domain  provides complementary information so we carried out AC susceptibility measurements, which also extend to higher frequencies and temperatures (Fig. 2b,c, and Supplementary Information Fig. S2). Surprisingly, two distinct characteristic frequencies are apparent for $T=1.5$~K (Fig. 2b). This contrasts with the well-defined single-mode response that we observe at low temperatures and reported in literature~\cite{quilliam2011dynamics,eyvazov2018common}. As $\chi(f)$ at low temperature can be described by the empirical Cole-Davidson form (see Methods and Supplementary Information Fig. S3)
\begin{equation}
\chi(\omega=2\pi f)=\chi'-i\chi''=\frac{\chi_0}{(1+i\omega\tau)^{\beta}},(0<\beta\leq1), 
\label{C-D susceptibility}
\end{equation}
we use a superposition of two such response functions to describe the bi-modal spectrum at $T=1.5$~K (Fig. 2b):
\begin{equation}
\chi(\omega=2\pi f) = \chi_l + \chi_h = \frac{\chi_{0l}}{(1+i\omega\tau_l)^{\beta_l}}+\frac{\chi_{0h}}{(1+i\omega\tau_h)^{\beta_h}},(0<\beta_l,\beta_h\leq1).
\label{double susceptibility}
\end{equation}
Here $\chi_l$, $\chi_h$ refer to the response that dominates at low and high temperatures, respectively. $\chi_{0l}$ and $\chi_{0h}$ are the corresponding static susceptibilities and $\tau_{l,h}$ are the characteristic relaxation times. The numerical constants $\beta_{l,h}$ characterize the distribution of relaxation time scales. As in the time-domain stretched exponential function~\cite{palmer1984models}, $\beta=1$ represents Debye relaxation with a single time scale while $\beta<1$ describes an asymmetric spectral function with extra weight in a high-frequency tail. To stabilize the fitting analysis we fix $\beta_l$ and $\beta_h$ to be temperature-independent and determined them in a simultaneous global fit to all the susceptibility data, which yielded $\beta_l=0.73$ and $\beta_h=0.49$ (see Methods and  Supplementary Information Fig. S4). This suggests distinct microscopic characters for the $l$ and $h$ components of the relaxation response.

Fig. 3b also includes as open symbols data from two separate previous studies of the AC susceptibility of $\rm Ho_2Ti_2O_7$~\cite{quilliam2011dynamics,eyvazov2018common} that would initially appear to be inconsistent with each other and with the present results.  We attribute the 3-4 decades longer time scales in the current work (Fig. 3b) to our highly-stoichiometric traveling-solvent-floating-zone grown crystals~\cite{ghasemi2018pyrochlore}. Oxygen vacancies and $\rm Ho^{3+}$ stuffing associated with conventional floating-zone grown samples may offer nucleation centers for monopole creation~\cite{sala2014vacancy} and effective transverse fields on the non-Kramers $\rm Ho^{3+}$ ions that increase the relaxation rate~\cite{ehlers2008dynamic}. Though two modes were not previously resolved in $\rm Ho_2Ti_2O_7$, there were indications of a faster relaxation process in the previously published lineshape analysis for $\chi(f)$. Specifically,  Ref.~\cite{quilliam2011dynamics} documents the abrupt appearance of an asymmetric lineshape upon heating, that is also apparent in a single mode analysis of our data (Fig. 3d).

The $\bf Q_\perp$ independent nature of the field dependent diffuse magnetic scattering (Fig. 1e,h), indicates that magnetic relaxation in the low $T$ regime for $T<1.25$~K is driven by the motion of thermally-activated monopoles along the applied field direction (Supplementary Information, Note 1 and Fig. S1). Defining the monopole conductivity through the magnetic version of “Ohm’s law”:  ${\bf J}_{\rm m} =\sigma_{\rm m} {\bf H}$, the magnetization relaxation time $\tau=\sigma_{\rm m}^{-1}$. Charge relaxation in a semiconductor where $\tau_{\rm e}=\epsilon_0/\sigma$ provides the electrical analogue. In this framework, the rise in the relaxation time upon cooling (Fig. 3b) simply indicates the increasing monopole resistivity anticipated for a monopole ``semi-conductor".  Within the dumbbell model~\cite{castelnovo2011debye}, the monopole conductivity can be written $\sigma_{\rm m}(T)=\mu(T)\mu_0Q_{\rm m}^2n(T) $, where $\mu_0$ is the vacuum permeability and $\mu(T)$, $Q_{\rm m}$, and $n(T)$ are the monopole mobility, charge, and number density respectively. Much as the resistivity of a semiconductors, the magnetization relaxation times $\tau_l(T)$ and $\tau_h(T)$ increases upon cooling in a manner that can be described by the Arrhenius law (Eqn.\ref{Arrhenius_law}). However, the barrier height $\Delta_l=15.8$~K (Table 1) exceeds the energy of a free monopole $\Delta_\text{mono}$, which is estimated to be $5.7$~K for $\rm Ho_2Ti_2O_7$ and $4.35$~K for $\rm Dy_2Ti_2O_7$ in the low temperature limit~\cite{castelnovo2011debye, jaubert2011magnetic}. This indicates the monopole mobility $\mu(T)$ decreases with $T$ in proportion to a power of the monopole density: $\mu\propto n^{\eta}$ with $\eta_l\simeq1.8$ for high-quality $\rm Ho_2Ti_2O_7$, and $\eta\simeq1.2$ for $\rm Dy_2Ti_2O_7$ (Table 1)~\cite{castelnovo2011debye,revell2013evidence}. Monopole motion is ultimately associated with quantum tunneling of the flippable spins adjacent to monopoles with a tunneling rate that may be enhanced by transverse magnetic fields from more distant monopoles~\cite{tomasello2019correlated}. Different values of $\eta$ for $\rm Ho_2Ti_2O_7$ and $\rm Dy_2Ti_2O_7$ are not unexpected due to the distinct non-Kramers vs. Kramers properties of the ground state doublet. Hyperfine coupling to nuclear spins may also play a role and is different for $\rm Dy^{3+}$ and $\rm Ho^{3+}$~\cite{paulsen2019nuclear}. None of these effects are captured by Monte Carlo simulation because they impact the metropolis time. Neglecting the $T$-dependence of the metropolis time, Monte Carlo simulation yields $\mu\propto 1/T$~\cite{castelnovo2011debye}, which is inconsistent with the data. 

The width of the peak in $\chi''(\log(f))$ is a measure of the distribution of relaxation times. In glass-forming systems the relaxation time distribution typically broadens upon cooling~\cite{binder1986spin}. A concomitant decrease in $\beta$ is an alternate indicator of glassy heterogeneous relaxation as in proton glasses which display logarithmic time dependence ($\beta\rightarrow 0$)~\cite{feng2006quantum}. In $\rm Ho_2Ti_2O_7$ on the other hand, the width of the peak in $\chi''(\log(f))$ actually narrows upon cooling. Fig. 3d shows that the HWHM of the peak in $\chi''(\log(f))$ approaches the limit for a single characteristic relaxation time while $\tau_l$ increases by six orders of magnitude and $\beta$ remains close to 1 (Fig. 3c). This observation of Debye relaxation over a $10^5$~s timescale in the low $T$ limit is unique to our knowledge and indicative of the independent motion of individual monopoles at low density through a homogeneous spin ice vacuum. This interpretation is supported by the deviation of $\tau_l$ from Arrhenius fitting in a notably flat fashion (Fig. 3b), which indicates loss of homogeneity as the ``ice rule" is violated at elevated temperatures. A distinct relaxation mode thus emerges (Fig. 3a), which will be discussed shortly. 

It is instructive to compare the low$-T$ Arrhenius parameters for $\rm Ho_2Ti_2O_7$ crystals of different qualities (Table 1). As expected the monopole “resistivity” $\tau$ is larger in higher quality samples. In fact, while the Arrhenius barrier height $\Delta_l$ is 50\% larger, the asymptotic relaxation time $\tau_{0l}$ is a factor of three smaller (Table 1). This indicates monopole motion at low $T$ is significantly modified in the more disordered samples. We speculate that stuffing (interstitial Ho spins) as well as other structural disorders~\cite{shafieizadeh2018superdislocations} may enhance spin tunneling whence the monopole mobility. 

We now turn to the $T>1.25$~K regime where a new mode of relaxation becomes dominant (Fig. 3a). Monopoles subject to competing hopping processes with time scales given by $\tau_l(T)$ and $\tau_h(T)$ respectively would result in a branching ratio $\chi_{0l}/\chi_{0h}=(\tau_{0l}/\tau_{0h})\exp [(\Delta_l-\Delta_h)/T]$, implying a cross over where $\chi_{0l}\approx\chi_{0h}$ for  $T=(\Delta_h-\Delta_l)/\ln ({\tau_{0l}/\tau_{0h}})=0.69$~K, which is inconsistent with the observed cross-over temperature of $T=1.25$~K (Fig. 3a). On the other hand, associating the high mode with a minimum energy cost $\Delta E $ would lead to $\chi_{0l}/\chi_{0h}=C\exp(\Delta E/T)$, which best fits the data in Fig. 3a with a numerical constant $C=3.4(6)\times10^{-4}$ and $\Delta E=9.7(9)$~K (Supplementary Information Fig. S5a). Both the cross-over temperature $T>>0.69$~K and the numerical factor $C<<1$ indicate the high$-T$ mode arises from a magnetizable entity that is distinct from isolated monopoles. The activation energy $\Delta E=9.7$~K is close to the energy cost to flip a spin out of the spin ice manifold~\cite{jaubert2011magnetic,castelnovo2011debye} so that a natural candidate is the Bjerrum monopole anti-monopole pair~\cite{giblin2011creation, castelnovo2011debye}. As its characteristic dimension $l_{\rm B}(T)=(\mu_0Q_{\rm m}^2/8k_{\rm B}T)$ decreases on warming, the Bjerrum pair becomes a spin-flip relative to the spin-ice manifold when $l_{\rm B}(T)$ approaches the diamond lattice constant $a_{\rm d}$. The corresponding magnetic susceptibility is $\chi_{\rm B}(T)=(3\sqrt{3}/2)(\mu_z^2\rho_{\rm B}(T)/a_{\rm d}^3k_{\rm B}T)$, where  $\rho_{\rm B}$ is the pair density (Supplementary Information Note 2). Associating $\chi_{0h}(T)$ with $\chi_{\rm B}(T)$ we can infer $\rho_{\rm B}(T)$, which is  qualitatively consistent with the prediction from Debye-H\"uckel theory~\cite{castelnovo2011debye} though the experiment yields a sharper increase of $\rho_{\rm B}(T)$ with $T$ and saturation is achieved at lower temperature (Supplementary Information Fig. S5b). These differences between the model and data might be attributed to Bjerrum pairs trapped during cooling at any realistic rate~\cite{castelnovo2010thermal}. Nonetheless, the lower value of $\beta_h=0.49$ (Supplementary Information Fig. S4) indicates a broader spectrum of relaxation times with an asymptotic time scale $\tau_{0h}$ that increases by a factor of 20 in disordered crystal (Table 1). Both features consistently point to spin-like high$-T$ relaxation with a strongly disorder-dependent relaxation rate (Fig. 3a).

While monopolar and dipolar entities in spin ice are indistinguishable in  thermodynamic observables such as heat capacity and static magnetization, we have shown they can be distinguished in the relaxation dynamics. It would appear then that $\rm Ho_2Ti_2O_7$ never really freezes but upon cooling undergoes a cross-over from dipolar to monopolar relaxation. This cross-over is not apparent in the reported work on $\rm Dy_2Ti_2O_7$~\cite{matsuhira2011spin,yaraskavitch2012spin} (Supplementary Information Fig. S6), which differs from $\rm Ho_2Ti_2O_7$ both in the nature of the single-ion spin and their exchange interactions. The approach to slow Debye relaxation in the low $T$ regime is inconsistent with conventional spin freezing but evidence of relaxation through motion of a diminishing density of monopoles through the spin ice manifold. That this regime is realized in high-quality single crystals is encouraging for the prospects of coherent quantum dynamics of monopoles~\cite{gingras2014quantum} in quantum siblings such as perhaps $\rm Ce_2Zr_2O_7$~\cite{gao2019experimental,gaudet2019quantum}.

\begin{methods}
\subsection{Samples and demagnetization factors}
All the $\rm Ho_2Ti_2O_7$ samples investigated in this work were cut from the same single crystal grown by the method of traveling solvent floating zone (TSFZ) technique~\cite{ghasemi2018pyrochlore}. In comparison, crystals investigated in Refs.~\cite{quilliam2011dynamics,eyvazov2018common} are from conventional floating zone growth, which typically manifests off-stoichiometry of $x>0.01$ in $\rm Ho_2(Ti_{2-\mathnormal x}Ho_{\mathnormal x})O_{7-\delta}$~\cite{ghasemi2018pyrochlore}. The dimensions of each sample and the demagnetization factors~\cite{sato1989simple, aharoni1998demagnetizing} employed in the data analysis are listed in the following table. 
\begin{center}
 \begin{tabular}{|c | c | c |} 
 \hline
 experimental methods & dimension (mm) ($l$ is along field direction) & demagnetization factor \\ 
 \hline
 neutron scattering & cylinder diameter~$ = 7.0$, $l = 8.0$ & 0.28~\cite{sato1989simple}  \\
 \hline
 SQUID & cross section~$=3.0\times0.53$, $l=5.0$ & 0.09~\cite{aharoni1998demagnetizing} \\
 \hline
 ACDR, ACMS & cross section~$=0.92\times0.64$, $l=2.0$ & 0.16~\cite{aharoni1998demagnetizing}  \\
 \hline
\end{tabular}
\end{center}
\subsection{Time-resolved neutron scattering~\cite{NISTdisclaimer}.}
With GE varnish at the interface, the $\rm Ho_2Ti_2O_7$ crystal was fastened to the top of a single crystal sapphire rod using teflon strips. The rod was mounted in a holder made of oxygen-free high thermal conductivity (OFHC) copper by GE varnish and black epoxy (STYCAST 2850FT). The copper holder was then threaded to the Helium-3 pot of the cryostat. Two RuO$_2$ thermometers (Lakeshore Rox-102A) were separately mounted on the sapphire rod and to the top of the $\rm Ho_2Ti_2O_7$ crystal, and read by the two channels of a Lakeshore 340 Temperature Controller. The magnetic field of hundred Oe was generated by an aluminum solenoid coil, centered around the sample and mounted around the inner vacuum chamber (IVC) made of silicon, while current of 1-2 Ampere was provided by a Dynatronix-DCR DPR20-15-30(XR) power supply with programmable output time sequences. Pump periods from $80$~ms to $200$~s were controlled by the internal clock of the power supply. Longer pump periods were externally controlled by a pulse generator (Berkeley Nucleonics Corporation Model 565). 

\noindent
Elastic neutron scattering was performed with incident energy $E_{\text{i}}= 5$~meV on triple-axis spectrometers SPINS and MACS at NCNR for (002) diffraction and diffuse scattering, respectively. Time series of events from neutron detector(s) and the step change of turning on the magnetic field were recorded by a time stamper (General Electric, RS-DCS-107M4968), and then folded in a single pump period with $t=0$ defined as the moment of field on. While measuring diffuse scattering at MACS, diffraction angle A3 was oscillating continuously, leading to a time-dependent coverage of $\bf Q$ space, orthogonal to the periodic perturbation from the field. Thus a coverage function in the two dimensional A3-$\Delta t$ space was carried to properly calculate the scattering intensity and error bars in Counts/s (Fig. 1d-h, Supplementary Information Fig. S1). Data from MACS were analyzed by Mslice in DAVE~\cite{azuah2009dave} with the package of event mode updated on May 4, 2020. The time resolution of neutron probe can be estimated by the distribution of time taken for a neutron to be detected following an elastic scattering event at the sample, i.e. $\frac{dt}{t} = \sqrt{(\frac{dl}{l})^2+(\frac{dv}{v})^2}$, where the distance from sample to detector is $l\sim1.2$~m with $\frac{dl}{l}\simeq(\cos^{-1}(0.5^{\circ})-1)$ given the $1^{\circ}$ beam divergence for diffraction, and the speed of neutrons with incident energy of $E_\text{i} = 5\pm0.2$~meV is $v=977$~m/s with $\frac{dv}{v}=\frac{dE_\text{i}}{2E_\text{i}}\sim0.02$. Therefore, $dt\approx24$~\textmu s in our experimental configuration. On the pump side, the step change of current output from the power supply has a width of $50$~\textmu s and $150$~\textmu s for the rising and falling edge, respectively, which sets an additional limit for time resolution. All combined, we estimate the time resolution of our pump-probe setup to be  $10^{-4}$~s.

\subsection{AC susceptibility measurements by SQUID and ACDR.}
AC susceptibility in absolute unit (emu Oe$^{-1}$ cm$^{-3}$) was measured by the commercial unit ACMS of Quantum Design PPMS from $1.9$ to $7$~K in the frequency range of $10-10^4$~Hz (Supplementary Information Fig. S2), which quantitatively anchored the lower-temperature measurements at $1-3.7$~K on the same sample by ACDR (Fig. 2c, Supplementary Information Fig. S2), a commercial unit of PPMS dilution fridge option, over the same frequency range. A SQUID-based magnetometer (iMAG 303 Multi-Channel SQUID System) with the low-pass filter of $5$~kHz was employed to measure AC susceptibility down to $f=1$~mHz in the dilution fridge (Fig. 2b), where the sample was glued on a silver plate with GE varnish and the drive and pick-up coils were mounted directly on the mixing chamber. A lock-in amplifier (Stanford Research, SR830) was employed to provide sinusoidal form of driving signal and pick up the in-phase and out-of-phase signals from the output of the SQUID control unit, which measures $\chi'(f)$ and $\chi''(f)$, respectively. The remnant uncompensated signal, generated by mismatched inductance of coils mounted on the mixing chamber, produced a temperature-independent background which was measured at $T=0.5$~K when the response from $\rm Ho_2Ti_2O_7$ is known to be too slow to generate any signal in the measured frequency range. The response function of the SQUID circuits was determined by comparing frequency scans at $T=4$~K that were separately measured by SQUID and ACMS, which meanwhile calibrated the conversion factor from millivolt to emu for the SQUID magnetometer to be $3.1\times10^{-7}$~emu/mV. All the presented AC susceptibility curves (Fig. 2b,c, Supplementary Information Fig. S2 and Fig. S3) were corrected for demagnetization effects using the equations given by Ref.~\cite{quilliam2011dynamics} for cgs units. Measurements from SQUID and ACDR match within 10\% as demonstrated by the $\chi(f)$ curves at $T=1.5$~K (Supplementary Information Fig. S2c), after their independent calibration by ACMS measurements in the separate ranges of overlap.

\subsection{Data analysis in time and frequency domains.}
All the fittings are performed using the Global Optimization Toolbox of Matlab, looking for convergence across sixty different starting points in the parameter space. AC susceptibility measured in frequency domain can be empirically described by Havriliak-Negami form of $\chi(\omega=2\pi f)=\frac{\chi_0}{(1+(i\omega\tau)^{\alpha})^{\beta}}(0<\alpha,\beta<1)$ \cite{rosa2015relaxation,kassner2015supercooled,eyvazov2018common}, which is reduced to Cole-Davidson form with $\alpha=1$ (Eqn.~\ref{C-D susceptibility}) and further reduced to Debye form if $\alpha=\beta=1$. We verfied that Cole-Davidson form is the minimal model to describe the measured $\chi(f)$ curves (Supplementary Information Fig. S3), consistent with previous investigation in $\rm Dy_2Ti_2O_7$, which has further argued $\beta\sim0.7-0.8$ to be the ideal case of $\beta=1$ modified by boudary effects~\cite{revell2013evidence}.
The forms given by Eqn.~\ref{C-D susceptibility} and Eqn.~\ref{double susceptibility} have been respectively used for analysis of single mode and two modes, while the simultaneous fitting of real and imaginary parts of $\chi(f)$ automatically respects the Kramers-Kronig relationship. 

\noindent
We first describe the single-mode analysis whose ultimate failure at high temperature is clearly demonstrated in Supplementary Information Fig. S3. For susceptibility $\chi(f)$ described by Eqn.~\ref{C-D susceptibility}, its integrated Fourier transformation is accordingly employed to describe the time-domain relaxation function $M(t)$, in response to a step change of magnetic field, measured by neutron scattering, which is
\begin{equation}\label{C-D relaxation_original}
M(t)=M_0[1-(t/\tau)^\beta E^{\beta}_{1,\beta+1}(-t/\tau)],
\end{equation}
where $M_0$ is the magnetization at equilibrium, and $E^{\beta}_{1,\beta+1}$ is three-parameter Mittag-Leffler function~\cite{rosa2015relaxation, haubold2011mittag}, which was numerically
evaluated~\cite{garrappa2015numerical}. Practically for demagnetization correction, we employed the numerically generated $M(t)$ array from
\begin{equation}\label{C-D relaxation}
M(t)=\int_{-\infty}^{t} \chi(t-t')[H_0-4\pi DM(t')] dt'
\end{equation}
where
\begin{equation}\label{Mittag-Leffler}
\chi(t)=-\frac{\chi_0}{\tau}(t/\tau)^{\beta-1}E^{\beta}_{1,\beta}(-t/\tau)
\end{equation}
represents the response to a $\delta$-function impulse, which is a direct Fourier transform of $\chi(f)$ given by Eqn.~\ref{C-D susceptibility}~\cite{rosa2015relaxation, haubold2011mittag}. At a given temperature, $\chi_0$ was extracted from the interpolation or extrapolation from susceptibility measurements, $H_0=100$~Oe is the external magnetic field, and $D$ is the demagnetization factor. 

\noindent
Neutron scattering intensity at $Q=0$ positions connects to $M(t)$ through
\begin{equation}\label{Time-resolved neutron}
I(t)=I_0+AM^2(t)
\end{equation}
where $I_0$ is nuclear scattering background, and $A$ is a numerical factor determined by the scattering geometry and cross section. When equilibrium is reached at $t>>\tau$, the fully developed magnetic moment $M_\text{eq}=\frac{\chi_0H_0}{1+4\pi D\chi_0}$ leads to neutron scattering intensity $I_\text{eq}=I_0+AM_\text{eq}^2$. The time-dependent curves presented in Fig. 2a are scaled as $(I(t)-I_0)/(I_\text{eq}-I_0)$. We note that Eqn.~\ref{C-D relaxation_original} employing Mittag-Leffler function is analogous to the stretched exponential function $\exp(-(t/\tilde{\tau})^{\tilde{\beta}})$~\cite{kassner2015supercooled, alvarez1991relationship}, but with the advantage that $\tau$ and $\beta$ have exactly identical physical meaning with Eqn.~\ref{C-D susceptibility}, allowing a consistent notation with frequency-domain measurements (Fig. 3b,c). Such a single-mode formalism leads to temperature-dependence of $\beta$ above $0.9$~K and discrepancy between neutron and susceptibility results at high temperatures. This can be accounted by the modified lineshape of $\chi(f)$ from the high mode, which fails the single-mode description differently in the time- and frequency-domains due to the respectively linear and logarithmic sampling. In addition, the linear sampling in the time domain, along with the time-varying statistical quality of neutron scattering data, makes it not accurate to identify two modes in the neutron measured $M(t)$. 

\noindent
When applying the two-mode model (Eqn.~\ref{double susceptibility}) to $\chi(f)$ measured by SQUID measurement (Fig. 2b), fitting at temperatures above $T=1.14$~K could converge without any constraints or particular choice of starting points. However, when the two characteristic frequency values get closer and the spectral weight of the high mode gets weaker at lower temperature, convergence becomes ambiguous. In particular, spectral weight at high frequency can be partly accounted by $\beta_l<1$, a direct consequence of the Cole-Davidson functional form (Eqn.~\ref{C-D susceptibility}), which complicates the determination of $\chi_{0h}$, $\beta_h$ and $\tau_h$. In the following, we detail and legitimate the constraints employed in the fitting.

\noindent
We noticed that $\beta_l$ of the low mode is essentially the temperature-independent $\beta$ below $T\sim0.9$~K in the single-mode description (Fig. 3c), while $\beta_h$ was found to fluctuate within $0.48\pm0.03$ with no discernible temperature dependence at $T>1.14$~K (Supplementary Information Fig. S4a) when two modes can be unambiguously determined. Therefore, we applied a global optimization to reduce the redundant freedom and determined $\beta_l=0.728\pm0.004$ and $\beta_h=0.492\pm0.008$ for all temperatures (Supplementary Information Fig. S4b,c). We further conceived the best guess of $\tau_h$ to be what is predicted by Arrhenius law as extrapolated from the high temperature fitting, but only used it as the starting point of the fitting process and allowed it to vary. This was the only additional condition compared with fitting at temperature above $1.14$~K. Fitting results of $\chi_0$ and $\tau$ with the above conditions are presented in Fig. 3a,b. The nascent emergence of the high mode around $T\simeq0.9$~K captured by $\chi_{0h}$ (Fig. 3a) is consistent with signatures from the lineshape (Fig. 3c,d). 

\noindent
Finally, measurements from ACDR are fit solely by the high mode $\chi_h$, since spectral weight of the low mode falls beyond the frequency window of the instrument ($10 - 10^4$~Hz). The limited frequency range in addition affects the overall fitting quality due to the tails of the low mode creating a background shape that escapes accurate modeling (Fig. 2c, Supplementary Information Fig. S2). 

\noindent
The $1\sigma$ value for all the fitting parameters were determined by the boundary of fitting $\chi^2$ normalized by the minimum $\chi^2_\text{min}$ to be $1+1/N_d$, where $N_d$ is the total number of data points subtracted by the number of free parameters.

\noindent
\end{methods}

\bibliography{main}

\begin{thebibliography}{10}
\bibitem{NISTdisclaimer}
\bibinfo{title}{Certain commercial equipment, instruments, materials, or
  software are identified in this section to foster understanding. {S}uch
  identification does not imply recommendation or endorsement by the {National
  Institute of Standards and Technology}, nor does it imply that the materials
  or equipment identified are necessarily the best available for the purpose.}
\end{thebibliography}

\begin{addendum}
 \item We are grateful to J. Ziegler, A. Malone, Wangchun Chen, Guangyong Xu, Y. Hernandez, Qiang Ye, T. Dax and Y. Vekhov for the tremendous help during time-resolved neutron scattering experiments at NCNR. We acknowledge P. Holdsworth, R. Moessner, C. Castelnovo and O. Tchernyshyov for helpful discussions. This work at Johns Hopkins University was supported as part of the Institute for Quantum Matter, an Energy Frontier Research Center funded by the U.S. Department of Energy, Office of Science, Basic Energy Sciences under Award No. DE-SC0019331. Development of the time-resolved neutron scattering methods was supported by the Gordon and Betty Moore foundation's EPiQS Initiative under GBMF9456. Access to MACS was provided by the Center for High Resolution Neutron Scattering, a partnership between the National Institute of Standards and Technology and the National Science Foundation under Agreement No. DMR-1508249. The SQUID measurement at University of Tokyo was partially supported by CREST (JPMJCR18T3), Japan Science and Technology Agency (JST),  and by Grants-in-Aid for Scientiﬁc Research from JSPS (16H06345, 18H03880, 19H00650). 
 \item[Author Contributions] Y.W and C.B designed the project. Y.W., T.R., J.K., N.M., Y.Q., J.A.R., S.G. and C.B. developed the time-resolved neutron scattering technique and performed the experiments. Y.W., Y.K. and S.N. performed the SQUID measurements. Y.W. and T.H. did the ACDR and ACMS measurements. S.M.K. provided single crystals $\rm Ho_2Ti_2O_7$. Y.W., T.R. and C.B. analyzed data. Y.W. and C.B. prepared the manuscript, with input from all authors.
 \item[Competing Interests] The authors declare that they have no competing financial interests.
 \item[Data availability] The data that support the findings of this study are available from the corresponding authors.
\end{addendum}

\begin{figure}
    \centering
    \includegraphics[width=16.5cm]{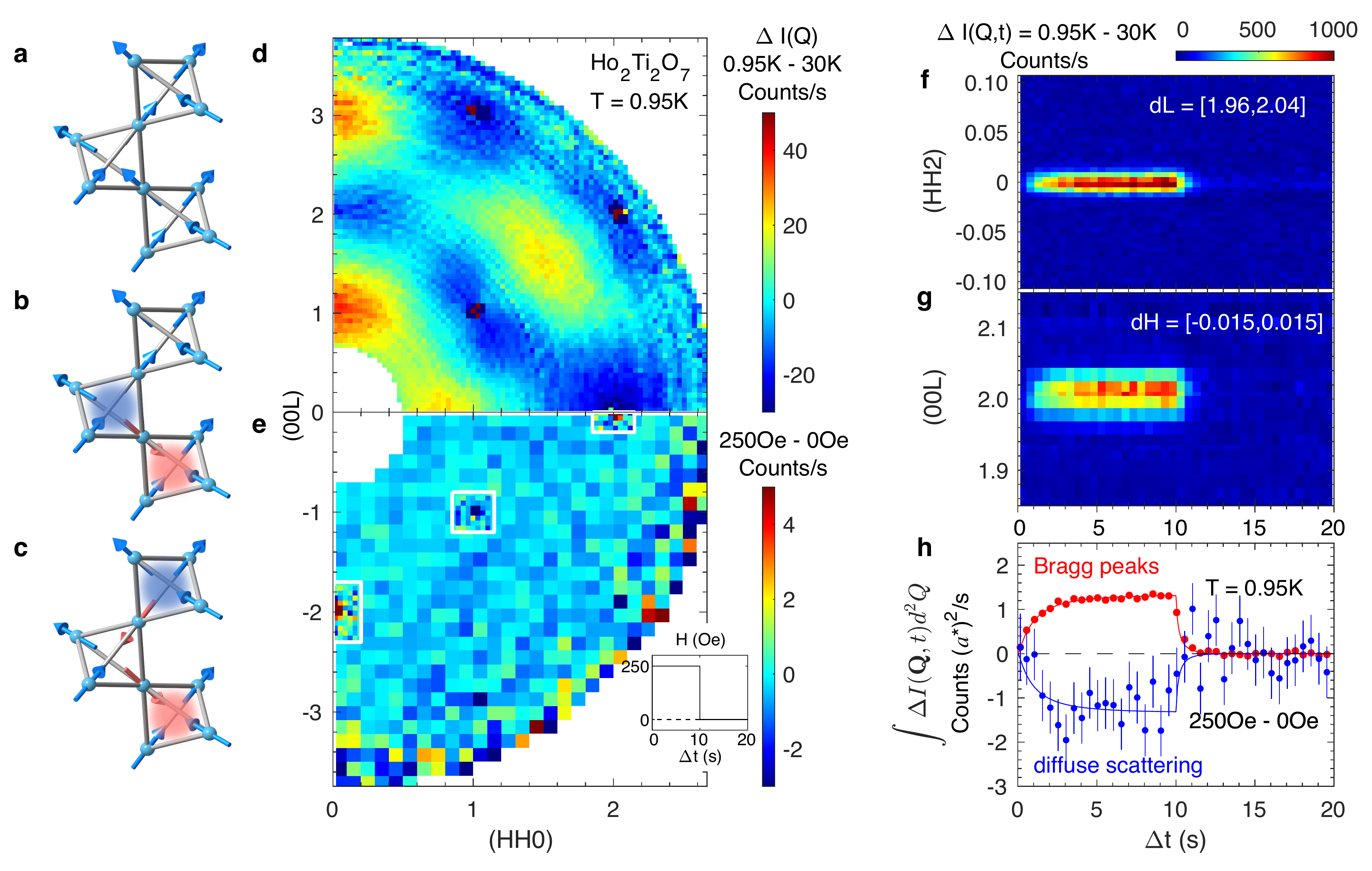}
    \caption{\textbf{Spin ice state and time-resolved neutron diffuse spectrum of $\mathbf{Ho_2Ti_2O_7}$.} \textbf{a-c,} Spin ice and magnetic monopoles on pyrochlore lattice. The spin ice state is constituted of corner-sharing tetrahedron with two-in-two-out spin configurations (\textbf{a}). A spin-flip excitation out of the ground state (\textbf{a} to \textbf{b}, red arrow) creates a pair of “three-in-one-out” (red cloud) and ``three-out-one-in" (blue cloud) defects. No further violation of the ``ice rule" takes place for successive spin flips which can be effectively described by the hopping of magnetic monopoles (\textbf{b} to \textbf{c}). Annihilation could happen when monopoles with opposite polarities conincide (\textbf{b} to \textbf{a}, reverse red arrow). \textbf{d-h,} Elastic diffuse scattering of $\rm Ho_2Ti_2O_7$ in $(HHL)$ plane at $T=0.95$~K probed by $5$ meV neutrons while pumped by a magnetic field of $250$~Oe along $\langle1\bar{1}0\rangle$ direction, which was periodically turned on and off for $10$~s each (\textbf{e} inset, $\Delta t$ is in reference to the closest step change of turning on the field). While zero-field diffuse scattering (\textbf{d}) demonstrates spin ice ground state, the field-induced difference (\textbf{e}) concentrates at ${\bf Q}=0$ positions, zoomed in with finer pixel sizes as surrounded by white frames. Magnetic signal is most clearly demonstrated at (002) and (220) where nuclear scattering is forbidden or weak. At the strong nuclear peak (111), the apparent decrease in intensity is only 1\% of the nuclear scattering, which can be well accounted by extinction release effect under field. Time-dependent spectrum in the vicinity of (002) is unfolded along $(HH0)$ (\textbf{f}) and $(00L)$ (\textbf{g}) directions. Integrated intensity in $(HHL)$ plane $\int{\Delta I({\bf Q},t)d^2Q}$ (\textbf{h}) documents opposite time-dependence at Bragg peaks (red), as sum of (002) and (220), and across the first quadrant excluding ${\bf Q}=0$ positions (blue). The red solid line represents best fit of Bragg intensity using Eqn.~\ref{Time-resolved neutron} (Methods), with the blue line its mirror image relative to the $x$-axis to clearly demonstrate the scattering sum-rule. A time constant of $\tau=3.6$~s is extracted after demagnetization correction (Methods). For all five panels (\textbf{d}-\textbf{h}), zero-field and $250$~Oe conditions refer to time windows of $12-20$~s and $2-10$~s respectively (\textbf{e} inset). Error bars represent 1$\sigma$ s.d. of counting statistics.}
    \label{fig:NeutronDiffuse}
\end{figure}

\begin{figure}
    \centering
    \includegraphics[width=16cm]{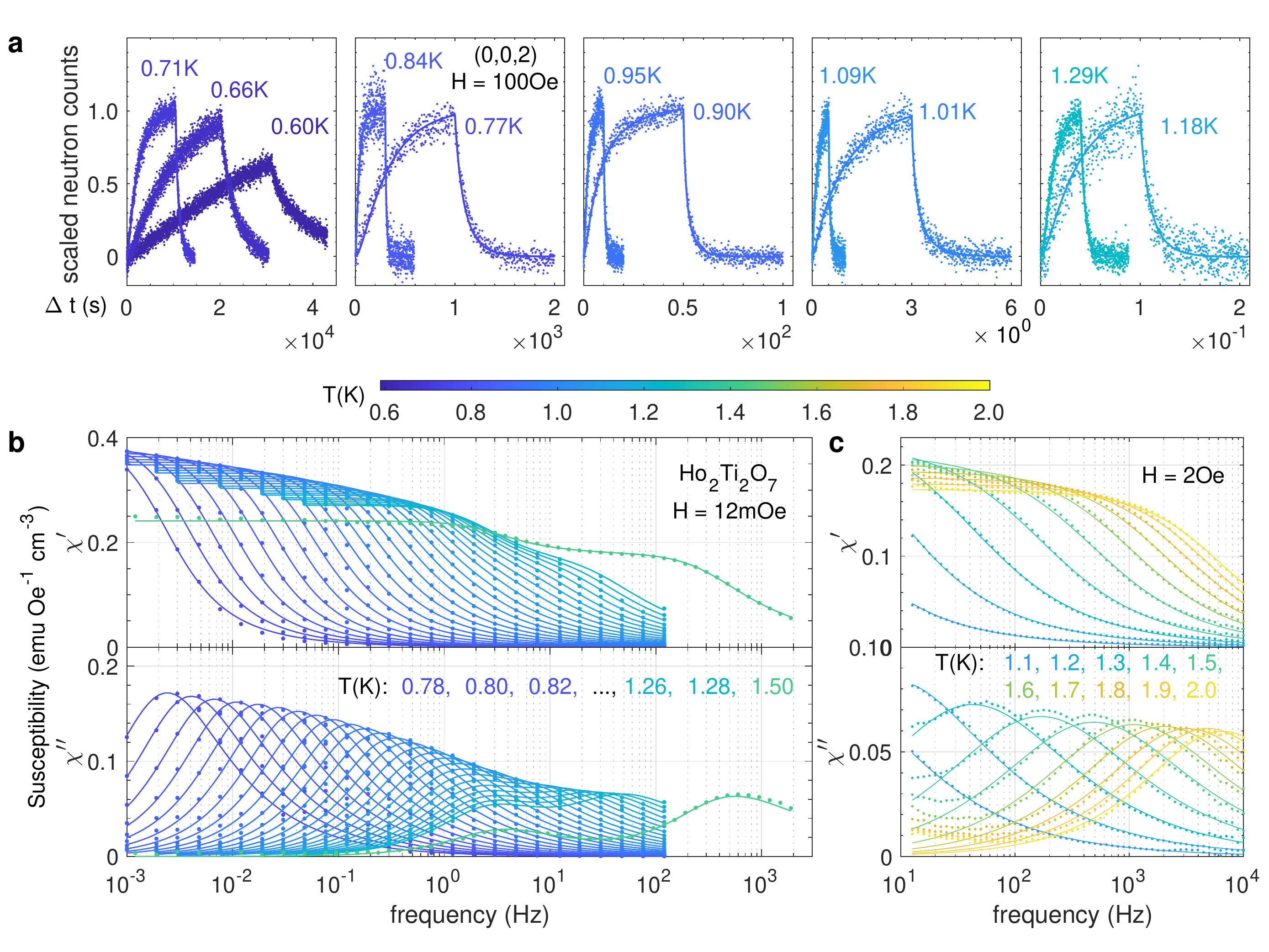}
    \caption{\textbf{Time-dependent neutron diffraction and frequency-dependent susceptibility, spanning over nine decades of timescales.} \textbf{a,} Time-dependent diffraction intensity of (002) under periodic perturbation from $H=100$~Oe at $T=0.6-1.3$~K, scaled to the expected intensity at equilibrium with field on (Methods). In each curve, the rising and dropping branches represent responses to step changes of turning on and off magnetic field respectively (Fig. 1\textbf{e} inset), with the pump period varying from the order of $10^{-2}$ to $10^4$~s. Solid lines represent best fits using Eqn.~\ref{Time-resolved neutron} (Methods). \textbf{b, c,} AC susceptibility $\chi(f)$ measured by a SQUID magnetometer in a dilution fridge down to $1$~mHz (\textbf{b}, Methods) and commercial Quantum Design ACDR susceptometer down to $10$~Hz (\textbf{c}, Methods). Real ($\chi'$) and imaginary ($\chi''$) parts of $\chi(f)$ are presented in top and bottom panels, respectively. SQUID measurements are limited by temperature stabilization between $T=1.3-1.5$~K in a dilution fridge. ACDR measurements, on the other hand, are limited to frequency $f\geq10$~Hz. Correction of response function, demagnetization effects and calibration to absolute unit have been applied to all the presented curves (Methods). Only selective curves from ACDR measurements are displayed for clarity (all curves are presented in Supplementary Information Fig. S2). Solid lines in panel \textbf{b} represent best fits using two modes (Eqn.~\ref{double susceptibility}) with $\beta_l=0.73$ and $\beta_h=0.49$ (Methods). In panel (\textbf{c}), solid lines represent best fits with Havriliak-Negami form (Methods) for the best guides to the eye, while the discrepancy at low frequency clearly demonstrates the contribution from a distinct mode. Across all panels, color codes of temperatures are consistent. Magnetic fields are applied along $\langle1\bar{1}0\rangle$ direction.}
    \label{fig:Raw data}
\end{figure}

\begin{figure}
    \centering
    \includegraphics[width=16cm]{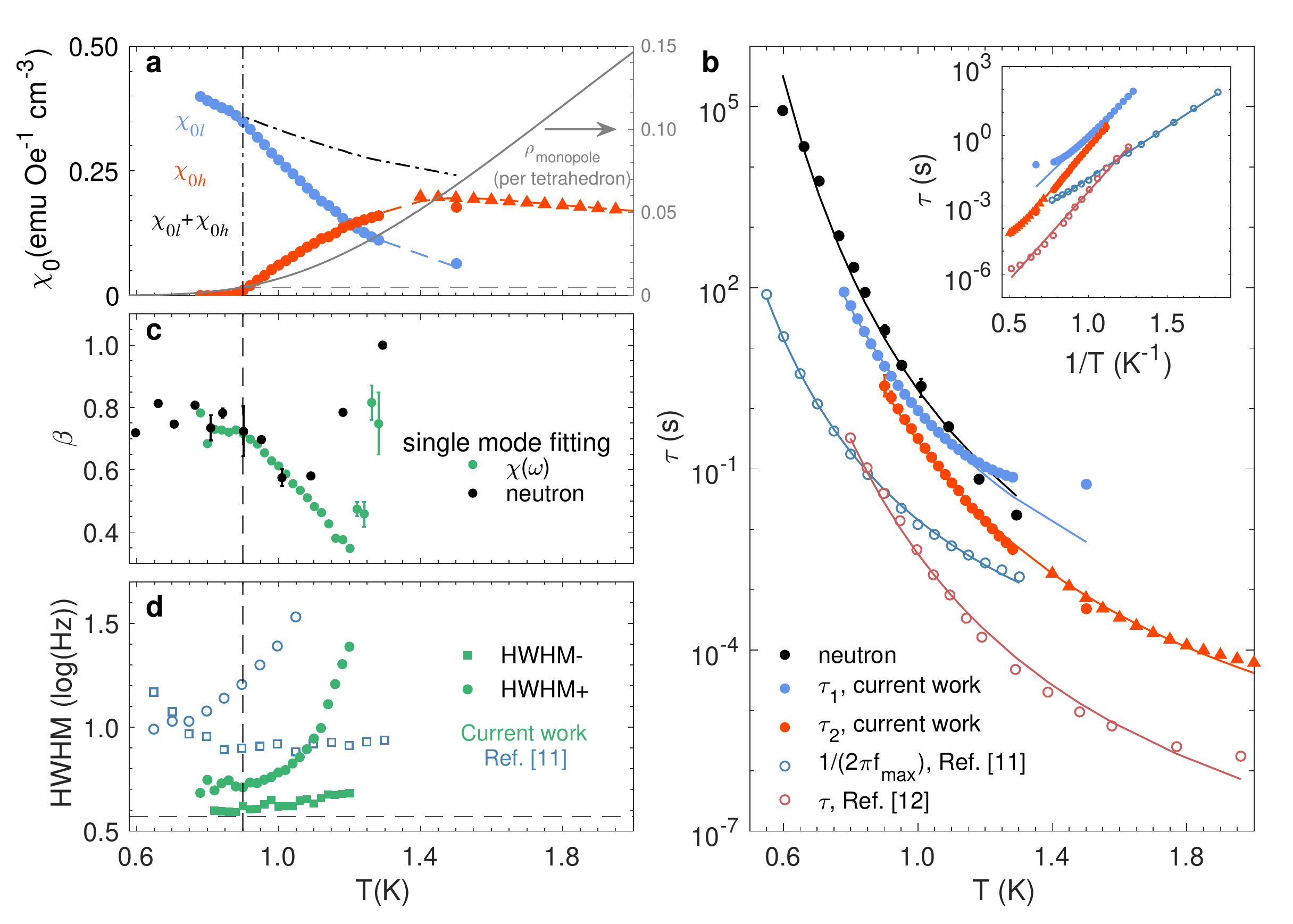}
    \caption{\textbf{Two relaxation modes in $\mathbf{Ho_2Ti_2O_7}$.} \textbf{a,b,} Temperature dependence of $\chi_{0}$ and $\tau$ for $\chi_l$ (light blue) and $\chi_h$ (red) in Eqn.~\ref{double susceptibility} with $\beta_l=0.73$ and $\beta_h=0.49$. Filled circles and triangles represent measurements by SQUID (Fig. 2\textbf{b}) and ACDR (Fig. 2\textbf{c}) respectively. Blue and red dash lines in \textbf{a} are guides to the eye, while the black line represents susceptibility at DC limit, $\chi_0=\chi_{0l}+\chi_{0h}$. The estimated monopole density (per tetrahedron) from Debye-H\"uckel theory~\cite{castelnovo2011debye} (Supplementary Information Note 2) is presented on the right $y$-axis of \textbf{a} (grey solid line), with the horizontal dashline located at $y=0.005$ indicating the estimated monopole density at $T\simeq0.9$~K. We present in \textbf{b} the time constants from neutron scattering (Fig. 2\textbf{a}), susceptibility measurements in current work (Fig. 2\textbf{b, c}), and published results in literature (Refs.~\cite{quilliam2011dynamics, eyvazov2018common}). Solid lines represent best fits of Arrhenius law with fitting parameters specified in Table 1. We note that $\tau$ measured by $M(t)$ is three times longer than the averaged $\tau_l$ and $\tau_h$, attributable to experimental artifacts such as thermal anchoring and imperfect demagnetization correction (Methods). A plot of $\tau-1/T$ is given in the inset to directly check the Arrhenius fitting. \textbf{c, d}, Lineshape analysis of neutron and susceptibility data below $T\sim1.3$~K using a single-mode measure. We documented the temperature evolution of exponent $\beta$ from susceptibility (Eqn.~\ref{C-D susceptibility}) and neutron scattering (Eqn.~\ref{Time-resolved neutron}, Methods), and the half-width-half-magnitude of  $\chi''(\log(f))$ on both sides of low- (HWHM-, square) and high-frequency (HWHM+, circle). The grey dash line in \textbf{d} indicates the expected HWHM of $0.57 \log$(Hz) for Debye relaxation. The temperature-independent asymmetry of HWHM+$>$HWHM- below $0.9$~K (\textbf{d}) directly manifests $\beta=0.73$ (\textbf{c}), while their dramatic temperature dependence starting from $0.9$~K signifies the onset of the high mode. The peak widths from Ref.\cite{quilliam2011dynamics} (\textbf{d}, open symbols) demonstrate $\sim50\%$ broadening effect and a lower onset temperature of the high mode presumably due to disorder effects.}
    \label{fig:Summary}
\end{figure}

\newpage

\begin{table}
\begin{center}
\begin{adjustwidth}{-0.25in}{-0.25in}
\begin{tabular}{ |>{\centering\arraybackslash}m{2.5cm}|>{\centering\arraybackslash}m{2.5cm}|>{\centering\arraybackslash}m{2.7cm}|>{\centering\arraybackslash}m{2cm}|>{\centering\arraybackslash}m{1.5cm}|>{\centering\arraybackslash}m{2cm}|>{\centering\arraybackslash}m{1.5cm}|} 
 \hline
 \multicolumn{7}{|l|}{$\rm Ho_2Ti_2O_7$, $J_{\text{eff}}\simeq1.8$~K}\\
 \hline
  & $\tau_{0l}$ (s) &  $\tau_{0h}$ (s) & $\Delta_{l}$ (K) & $\Delta_{l}/J_{\text{eff}}$ & $\Delta_{h}$ (K) & $\Delta_{h}/J_{\text{eff}}$  \\ 
 \hline
 current study & $2.3(8)\times10^{-7}$ & $4(1)\times10^{-9}$ & $15.3(3)$ & $8.5$  & $18.3(3)$ & $10.2$  \\
 \hline
 Refs~\cite{quilliam2011dynamics, eyvazov2018common}& $4.5(9)\times10^{-7}$& $9.5(7)\times10^{-11}$ & $10.4(2)$ & $5.8$ & $17.5(9)$ & $9.7$\\
 \hline
 \multicolumn{7}{|l|}{$\rm Dy_2Ti_2O_7$, $J_{\text{eff}}\simeq1.1$ K}\\
 \hline
  & \multicolumn{2}{|c|}{$\tau_0$ (s)} &\multicolumn{2}{|c|}{$\Delta$ (K)} &\multicolumn{2}{|c|}{$\Delta/J_{\text{eff}}$} \\ 
 \hline
 Ref.~\cite{yaraskavitch2012spin} & \multicolumn{2}{|c|}{$4.06\times10^{-7}$} & \multicolumn{2}{|c|}{$9.79$} & \multicolumn{2}{|c|}{$8.9$}  \\
 \hline
\end{tabular}
\end{adjustwidth}
\caption{\textbf{Arrhenius parameters for $\mathbf{Ho_2Ti_2O_7}$ and $\mathbf{Dy_2Ti_2O_7}$.} We employ Arrhenius law $\tau=\tau_0\exp(\Delta/T)$ (Eqn.~\ref{Arrhenius_law}) to fit temperature dependence of $\tau_l$ and $\tau_h$ in $\rm Ho_2Ti_2O_7$, both in current study and reported in literature~\cite{quilliam2011dynamics, eyvazov2018common}. The frequency and temperature range of the the instruments involved in Ref.~\cite{quilliam2011dynamics} and Ref.~\cite{eyvazov2018common} suggest that $\tau_l$ and $\tau_h$ have been respectively measured. Samples investigated in current work are from traveling-solvent-floating-zone growth~\cite{ghasemi2018pyrochlore}, while samples studied in Refs~\cite{quilliam2011dynamics,eyvazov2018common} are from traditional floating-zone growth (Methods). Arrhenius parameters for $\rm Dy_2Ti_2O_7$~\cite{yaraskavitch2012spin} and $J_{\text{eff}}$ values for both $\rm Ho_2Ti_2O_7$ and $\rm Dy_2Ti_2O_7$~\cite{melko2004monte, jaubert2011magnetic} are provided for comparison. Arrhenius fitting for $\tau(T)$ from neutron measurements gives $\Delta= 18(2)$~K and $\tau_0 = 4(8) \times 10^\text{-8}$~s, which are not listed for comparison due to the mixture of $\tau_l$ and $\tau_h$.}
\label{table:Arrhenius}
\end{center}
\end{table}

\end{document}


\usetagform{supplementary}
\title{\parbox{16cm}{Supplementary Information for: Monopolar and dipolar relaxation in spin ice Ho$_2$Ti$_2$O$_7$}}
\author{Yishu Wang$^{1,2,*}$, T. Reeder$^1$, Y. Karaki$^3$, J. Kindervater$^1$, T. Halloran$^1$, N. Maliszewskyj$^2$, Yiming Qiu$^2$, J. A. Rodriguez$^{2,4}$, S. Gladchenko$^2$, S. M. Koohpayeh$^{1,5}$, S. Nakatsuji$^{1,6,7,8}$ \& C. Broholm$^{1,2,5,*}$}

\maketitle

\begin{affiliations}
 \item Institute for Quantum Matter and Department of Physics and Astronomy, Johns Hopkins University, Baltimore, Maryland 21218, USA
 \item NIST Center for Neutron Research, National Institute of Standards and Technology, Gaithersburg, Maryland 20899, U.S.A
 \item Faculty of Education, University of the Ryukyus, Nishihara, Okinawa 903-0213, Japan
 \item Department of Materials Science and Engineering, University of Maryland, College Park, Maryland 20742, USA
 \item Department of Materials Science and Engineering, Johns Hopkins University, Baltimore, Maryland 21218, USA
 \item Institute for Solid State Physics, University of Tokyo, Kashiwa, Chiba 277-8581, Japan
 \item Department of Physics, University of Tokyo, Hongo, Bunkyo-ku, Tokyo 113-0033, Japan
 \item Trans-scale Quantum Science Institute, University of Tokyo, Hongo, Bunkyo-ku, Tokyo 113-0033, Japan
 \item[*] Correspondence: wangyishu@jhu.edu, broholm@jhu.edu
 \item[] \today
\end{affiliations}

\noindent
\textbf{Table of Contents}
\begin{enumerate}
\setlength\itemsep{0em}
    \item[Supplementary Note 1:]  ${\bf Q}$-dependence of field-induced diffuse scattering changes.
    \item[Supplementary Note 2:]  Estimates of monopole density, Bjerrum pair density and magnetic susceptibility from Debye-H\"uckel theory.    
    \item[Fig. S1:]  Cuts and integrated intensity across the reciprocal space in field-induced changes of neutron diffuse scattering.
    \item[Fig. S2:]  Supplementary data for AC susceptibility measurements.
    \item[Fig. S3:]  Cole-Davidson form as the minimal model to describe $\chi(f)$ and the breakdown of the single-mode description.
    \item[Fig. S4:]  Global determination of $\beta_l$ and $\beta_h$.
    \item[Fig. S5:]  Associating the high$-T$ mode with magnetizable dipoles.
    \item[Fig. S6:]  Comparative analysis of $\chi(f)$ lineshape between $\rm Ho_2Ti_2O_7$ and $\rm Dy_2Ti_2O_7$.
    \item[References] 
\end{enumerate}

\newpage
\section*{Supplementary Note 1: ${\bf Q}$-dependence of field-induced diffuse scattering changes.}

In this section, we estimate the ${\bf Q}$-dependence of field-induced changes in diffuse scattering, in particular its correlation with ${\cal S}({\bf Q})$. The unit of scattering intensity is counts/s, and that of integration over reciprocal space $\int d^2Q$ is $(\textit{a}^*)^2$. All the Bragg positions at Brillouin zone centers have been removed from the integration.

Spin ice correlation ${\cal S}({\bf Q})$ can be directly measured by diffuse scattering at low temperature, experimentally represented by $\Delta I_T({\bf Q}) = I_\text{0.95K}({\bf Q}) - I_\text{30K}({\bf Q})$ (Fig. 1d of the main text). For unity purpose, we define:
\begin{equation}
{\cal S}({\bf Q})=\frac{\Delta I_T({\bf Q})}{\int \Delta I_T({\bf Q})d^2Q}.  
\label{Eqn: S(q)}
\end{equation}
To distinguish the field-induced intensity that follows and does not follow ${\cal S}({\bf Q})$, we separate $\Delta I_H({\bf Q}) = I_\text{250 Oe}({\bf Q}) - I_\text{zero field}({\bf Q})$ (Fig. 1e) into two parts:
\begin{equation}
\Delta I_H({\bf Q}) = \Delta A\cdot {\cal S}({\bf Q}) +  \Delta B,  
\end{equation}
It follows that
\begin{equation}
\int \Delta I_H({\bf Q})d^2Q = \Delta A\cdot \int {\cal S}({\bf Q})d^2Q +  \Delta B \int d^2Q
\label{Eqn:int dI_H}
\end{equation}
\begin{equation}
\int \Delta I_H({\bf Q})\cdot {\cal S}({\bf Q})d^2Q = \Delta A\cdot \int {\cal S}^2({\bf Q})d^2Q +  \Delta B \int {\cal S}({\bf Q})d^2Q
\label{Eqn:int dI_H S(q)}
\end{equation}
where $\int {\cal S}^2({\bf Q})d^2Q=1$ as defined by Eqn. S\ref{Eqn: S(q)}, $\int {\cal S}({\bf Q})d^2Q=0$ if the integration covers all the energy and momentum transfer thus sum rule being ideal, while from the real data we get $\int {\cal S}({\bf Q})d^2Q=-0.061\pm0.002$. By solving the equation set formed by Eqn. S\ref{Eqn:int dI_H} and S\ref{Eqn:int dI_H S(q)}, and with experimental results $\int \Delta I_H({\bf Q})d^2Q=-1.28\pm0.15$, $\int \Delta I_H({\bf Q})\cdot {\cal S}({\bf Q})d^2Q=-0.01\pm0.04$, $\int d^2Q=4.4565$, we get $\Delta A=-0.03\pm0.04$ and  $\Delta B=-0.29\pm0.03$. Given the statistical error of $\Delta A$ and the fact $\Delta B\sim 10\Delta A$, we estimate that less than 10\% of the field-induced intensity follows ${\cal S}({\bf Q})$, if any.

$\Delta B$ estimates the total reduced intensity, which could either be ${\bf Q}$-independent, or ${\bf Q}$-dependent but not following ${\cal S}({\bf Q})$. To further demonstrate the reduced intensity being flat in ${\bf Q}$, we provide a few cuts across the reciprocal space and check the sum rule over different regions in Fig. S1.

\newpage
\section*{Supplementary Note 2: Estimates of monopole density, Bjerrum pair density and magnetic susceptibility from Debye-H\"uckel theory.}
Following the methods developed by Castelnovo \textit{et al} in Phys. Rev. B \textbf{84}, 144435 (Ref. 22 in the main text), we estimate the monopole density and Bjerrum pair density in $\rm Ho_2Ti_2O_7$ from Debye-H\"uckel (DH) theory. In the original paper, the authors presented quantitative results for $\rm Dy_2Ti_2O_7$ and provided the value of parameters for $\rm Ho_2Ti_2O_7$, which are used in the estimates here. We further calculate the susceptibility contributed by Bjerrum pairs and compare it with $\chi_{0h}$ measured in our experiment. 

The basic idea of DH theory is to introduce the magnetostatic energy into the mean-field free energy based on the dumbbell model, while minimizing free energy with respect to monopole density $\rho$ yields a self-consistent expression:
\begin{equation}
\rho = \frac{2\exp[-(\frac{\Delta}{T}-\frac{E_{\rm nn}}{2T}\frac{\alpha\sqrt{\rho}}{1+\alpha\sqrt{\rho}})]}{1+2\exp[-(\frac{\Delta}{T}-\frac{E_{\rm nn}}{2T}\frac{\alpha\sqrt{\rho}}{1+\alpha\sqrt{\rho}})]}.
\label{Eqn: rho_SC}
\end{equation}
$\Delta$ is the bare cost for an isolated monopole in the non-interacting picture. $E_{\rm nn}\equiv\mu_0Q_{\rm m}^2/4\pi a_{\rm d}k_{\rm B}$ is the Coulomb energy between two neighboring monopoles with magnetic charge $Q_{\rm m}\equiv2\mu/a_{\rm d}$, where $\mu$ is the spin magnetic moment and $a_{\rm d}$ is the diamond lattice constant. $\alpha(T)=\sqrt{\frac{3\sqrt{3}\pi E_{\rm nn}}{2T}}$ represents the temperature effect when mapping magnetostatic effect to the Debye approximation. The above equation can be numerically solved using a recursive approach: 
\begin{equation}
\rho_{0} = \rho_{\rm nn} = \frac{2\exp(-\frac{\Delta}{T})}{1+2\exp(-\frac{\Delta}{T})};\\
\rho_{l+1} = \frac{2\exp[-(\frac{\Delta}{T}-\frac{E_{\rm nn}}{2T}\frac{\alpha\sqrt{\rho_l}}{1+\alpha\sqrt{\rho_l}})]}{1+2\exp[-(\frac{\Delta}{T}-\frac{E_{\rm nn}}{2T}\frac{\alpha\sqrt{\rho_l}}{1+\alpha\sqrt{\rho_l}})]}
\label{Eqn:recursive}
\end{equation}
which could converge with acceptable accuracy in fewer than five steps. Therefore, $\rho=\rho_{l\rightarrow\infty}\simeq\rho_5$. The relevant parameters for $\rm Ho_2Ti_2O_7$ are $\Delta=5.79$~K, $\mu\simeq10\mu_{\rm B}$ ($\mu_{\rm B}$ refers to Bohr magneton), $a_{\rm d}=4.34$~\AA. The convergence value of $\rho(T)$ using these parameters for Eqn.~S\ref{Eqn:recursive} is presented in the main Fig. 3a on the right $y-$axis.

Similar analysis can be applied to the so-called Bjerrum pair, a bound state of monopole-antimopole within separation of a characteristic length $l_{\rm B}(T)=\mu_0Q_{\rm m}^2/8\pi k_{\rm B}T$. In $\rm Ho_2Ti_2O_7$, $l_{\rm B}$ is most conveniently estimated as 
\begin{equation} 
l_{\rm B}/a_{\rm d}\simeq \frac{1.54}{T[\rm K]}.
\label{Eqn:Bjerrum length}
\end{equation}
The density of Bjerrum pairs that are separated by $d$ diamond lattice constants, $\rho_d$, is given by
\begin{equation} 
\rho_d = \frac{v_d\exp(-\frac{2\Delta-E_d}{T})}{1+v_d\exp(-\frac{2\Delta-E_d}{T})},
\label{Eqn: bound pair density}
\end{equation}
where $v_d=2d^2$ is the number of configurations that the two monopoles in the pair can take, given the center of mass is fixed, and $E_d\sim E_{\rm nn}/d$ is the internal Coulomb energy of the pair. 

Eqns.~S\ref{Eqn: rho_SC}-S\ref{Eqn: bound pair density} are Eqns. (2.13), (2.14), (3.6) and (3.12) from the original paper by Castelnovo \textit{et al}.

Now we estimate the susceptibility contributed by Bjerrum pairs. According to Eqn.~S\ref{Eqn:Bjerrum length}, only pairs formed on nearest-neighbor diamond lattices, i.e. $d=1$, are relevant for the temperature range of the high mode resolved in our experiment ($T>0.9$~K). Therefore, we take $d=1$ in Eqn.~S\ref{Eqn: bound pair density} as Bjeruum pair density at equilibrium:
\begin{equation} 
\rho^{\rm eq}_{\rm B} \equiv \rho_{\rm B} (B=0) \simeq \rho_{d=1} = \frac{2\exp(-\frac{2\Delta-E_{\rm nn}}{T})}{1+2\exp(-\frac{2\Delta-E_{\rm nn}}{T})}.
\label{Eqn: NN bound pair density}
\end{equation}
In the presence of a magnetic field $B$, the bound pairs are divided into two classes: those with moments $\mu$ aligned and anti-aligned with $B$, whose density follows
\begin{equation} 
\rho_{\rm B}^{\pm} (B) = \frac{\exp(-\frac{2\Delta-E_{\rm nn}}{T})\exp(\pm\frac{2\mu_zB}{k_{\rm B}T})}{1+\exp(-\frac{2\Delta-E_{\rm nn}}{T})\exp(\frac{2\mu_zB}{k_{\rm B}T}))+\exp(-\frac{2\Delta-E_{\rm nn}}{T})\exp(-\frac{2\mu_zB}{k_{\rm B}T})}.
\label{Eqn: NN bound pair density in field}
\end{equation}
The $+$ and $-$ stand for the cases of moments parallel and anti-parallel with B, respectively, and $\mu_z=\mu_{\rm Ho}/2=5\mu_{\rm B}$ is the average projection on any given field direction of the $\langle111\rangle$-oriented Ho dipole moments. We note that a factor of 2 has been applied in the Zeeman energy since the magnetic dipole moment is contributed by spins on two tetrahedra. What is observable in experiments is 
\begin{equation} 
\Delta\rho_{\rm B}(B) = \rho_{\rm B}^{+} (B)-\rho_{\rm B}^{-}(B) \simeq \frac{\frac{4\mu_zB}{k_{\rm B}T}\cdot\exp(-\frac{2\Delta-E_{\rm nn}}{T})}{1+2\exp(-\frac{2\Delta-E_{\rm nn}}{T})}=\frac{2\mu_zB}{k_{\rm B}T}\rho^{\rm eq}_B, (\mu_zB<<k_{\rm B}T).
\label{Eqn: unbalanced bound pair density}
\end{equation}
Note that the density $\rho$ is estimated as number per tetrahedron, which is connected to the dimensionful monopole density $\rho_V$ by $\rho_V = 3\sqrt{3}\rho/(8a_{\rm d}^3)$. (As there are eight tetrahedra in one cubic unit cell with lattice constant $a_{\rm cubic} = \frac{4}{\sqrt{3}}a_{\rm d}$, the effective volume of a tetrahedron is $V=\frac{8a_{\rm d}^3}{3\sqrt{3}}$). Therefore, the volume susceptibility associated with bound pairs is:
\begin{equation} 
\chi_{\rm B} = \frac{2\mu_z}{B}\frac{\Delta\rho_{\rm B}(B)}{8a_{\rm d}^3/3\sqrt{3}}=\frac{3\sqrt{3}}{2}\frac{\mu_z^2}{k_{\rm B}a_{\rm d}^3}\frac{\rho^{\rm eq}_{\rm B}}{T}.
\label{Eqn: bound pair susceptibility}
\end{equation}
Numerically, $\mu_z=5\mu_{\rm B}=5\times(9.274\times10^{-21}\rm emu)$, $k_{\rm B}=1.3807\times10^{-16} {\rm emu}\cdot{\rm G}\cdot{\rm K^{-1}}$, and $a_{\rm d} = 4.34\times10^{-8}\rm cm$. Therefore,
\begin{equation} 
\chi_B \left[\frac{\rm emu}{\rm Oe\cdot cm^3}\right]\simeq0.49\left[\frac{\rm emu\cdot K}{\rm Oe\cdot cm^3/tetrahedron}\right] \frac{\rho^{\rm eq}_B [\rm{/tetrahedron}]}{T [\rm K]}.
\label{Eqn: bound pair susceptibility_numerical}
\end{equation}
We take the experimentally measured $\chi_{0h}$ (main Fig. 3a) as $\chi_B$ to estimate $\rho_B$, which is plotted in Fig. S5 together with what is calculated from Eqn.~S\ref{Eqn: NN bound pair density}.

\newpage
\begin{figure}
    \centering
    \includegraphics[width=13.5cm]{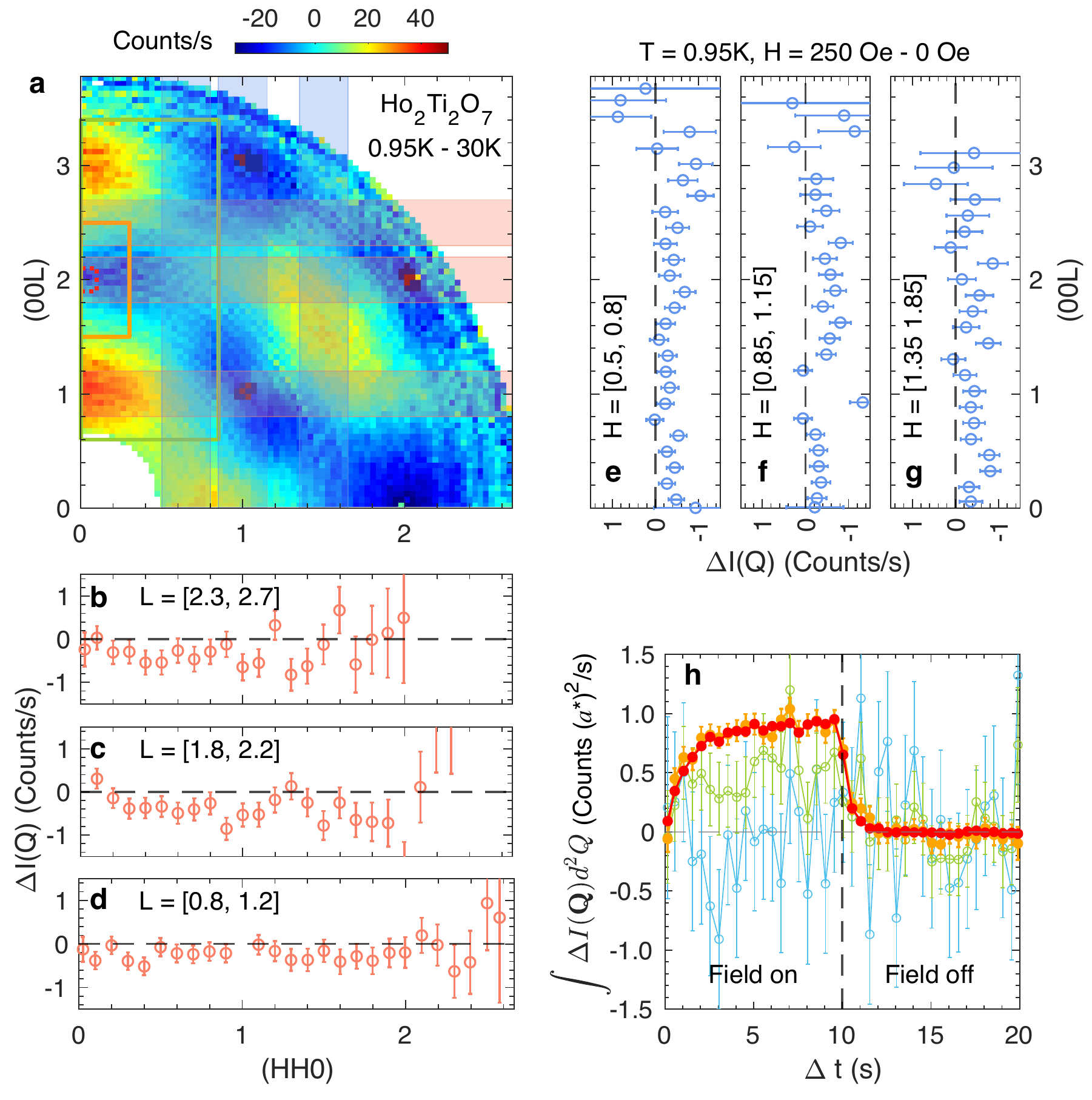}
    \caption{\textbf{Cuts and integrated intensity across the reciprocal space in field-induced changes of neutron diffuse scattering.} \textbf{a}, Zero-field neutron diffuse scattering at $T=0.95$~K with background at $T=30$~K subtracted, demonstrating the spin ice correlation. We patched the regions for cuts along $(HH0)$ (red, panels \textbf{b-d}) and $(00L)$ (blue, panels \textbf{e-g}) directions, and frame blocked the regions for integrated intensity (panel \textbf{h}). Note that the red dashed frame around the pinch point (002) was exaggerated for clarity, while the actual tiny range for integration is $H = [0, 0.03]$ and $L = [1.95, 2.05]$. Color codes between frames in \textbf{a} and traces in \textbf{h} are consistent, with an additional blue trace for the entire quadrant, excluding strong nuclear peaks at (111), (222) and (113). \textbf{b-g}, Cuts through $\bf Q$-space in the field-induced scattering at $H=250$~Oe, with zero-field spectrum subtracted as the background (main Fig.~1e). Negative intensity is consistently observed with no discernible $\bf Q$-modulation. \textbf{h}, Time-dependent integrated intensity with the magnetic field turned on and off periodically (main Fig.~1e inset). Scattering at Bragg peak (002) (red), with its vicinity included (orange), and in the broad area of a ``rod" along L-direction could not preserve the total spectral weight under field perturbation. The sum rule is observed only when the entire $\bf Q$ range is considered (blue). That is said, with field along $\langle1\bar{1}0\rangle$ direction, Bragg peaks are observed due to the polarization of spin chains out of the $(HHL)$ scattering plane, like spin dipoles, which then produce incoherent scattering suffusing entire $\bf Q$-space when field is turned off. Error bars represent 1$\sigma$ s.d. of counting statistics.}
    \label{supfig:ACDRraw}
\end{figure}

\newpage
\begin{figure}
    \centering
    \includegraphics[width=16.5cm]{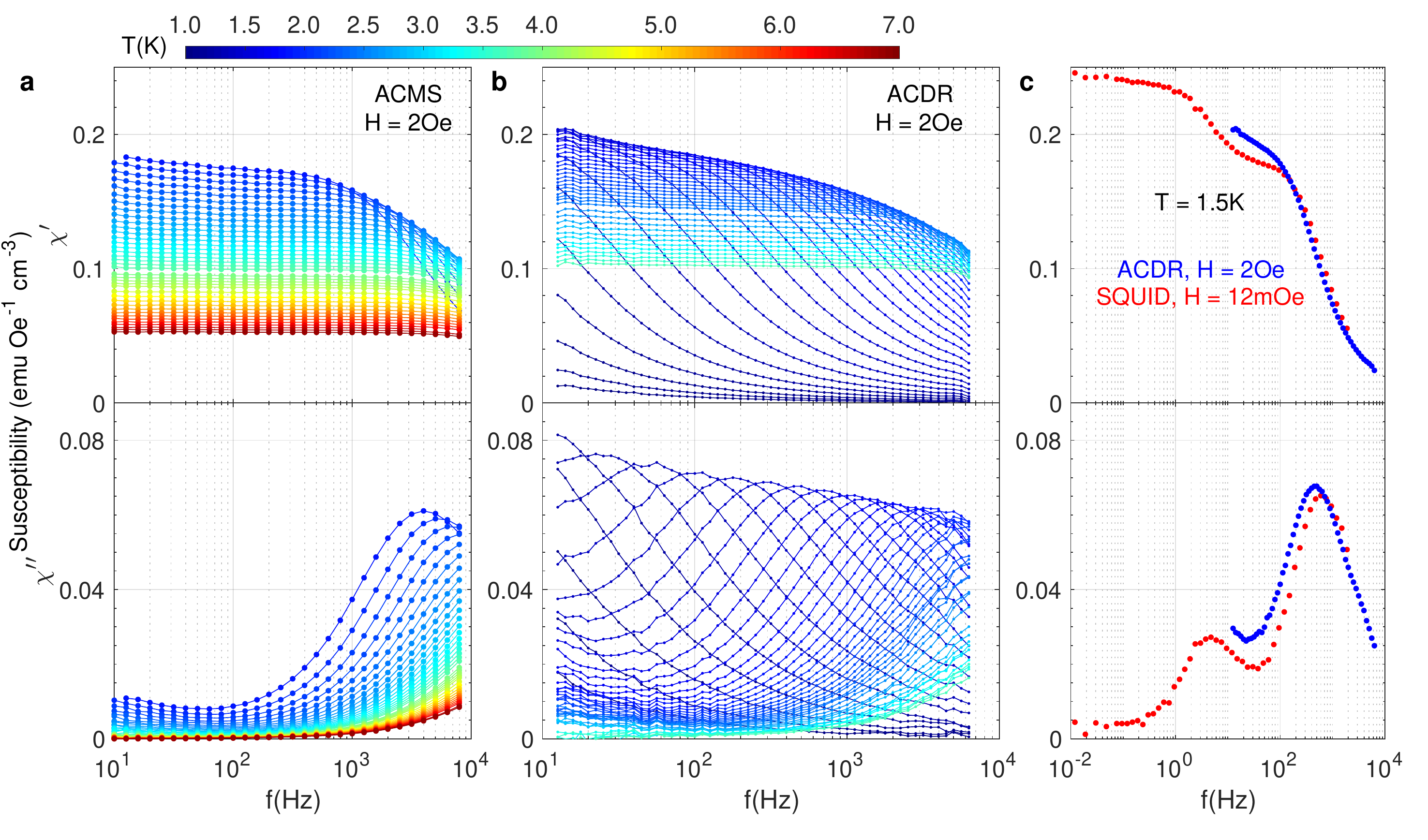}
    \caption{\textbf{Supplementary data for AC susceptibility measurements.} Real and imaginary parts of $\chi(f)$ are displayed in top and bottom panels, respectively. \textbf{a}, AC susceptibility measured by ACMS from $1.9$ to $7$~K, which provides values in the absolute unit (emu Oe$^{-1}$ cm$^{-3}$) (Methods)  \textbf{b}, AC susceptibility measured by ACDR (Methods) from $1$ to $3.7$~K on the same sample as measured in ACMS (panel \textbf{a}). Part of this data set is presented in the main Fig.~2c. The low mode lies below the frequency range of this instrument, but the apparent ``contamination" from the tailed low frequency mode can be recognized. In addition, for temperatures below $T=1.35$~K, even the high mode has a significant part of spectral weight lying below $10$~Hz, leading to high uncertainty in the fitting parameters $\chi_{0h}$ and $\tau_h$, which therefore are not included in the results presented the main Fig. 3a,b. Color codes for temperature are consistent across \textbf{a} and \textbf{b}. \textbf{c}, Susceptibility at $T=1.5$~K measured by ACDR (blue) and SQUID (red), after the independent calibration by their overlap with ACMS data.}
    \label{supfig:ACDRraw}
\end{figure}

\newpage
\begin{figure}
    \centering
    \includegraphics[width=17cm]{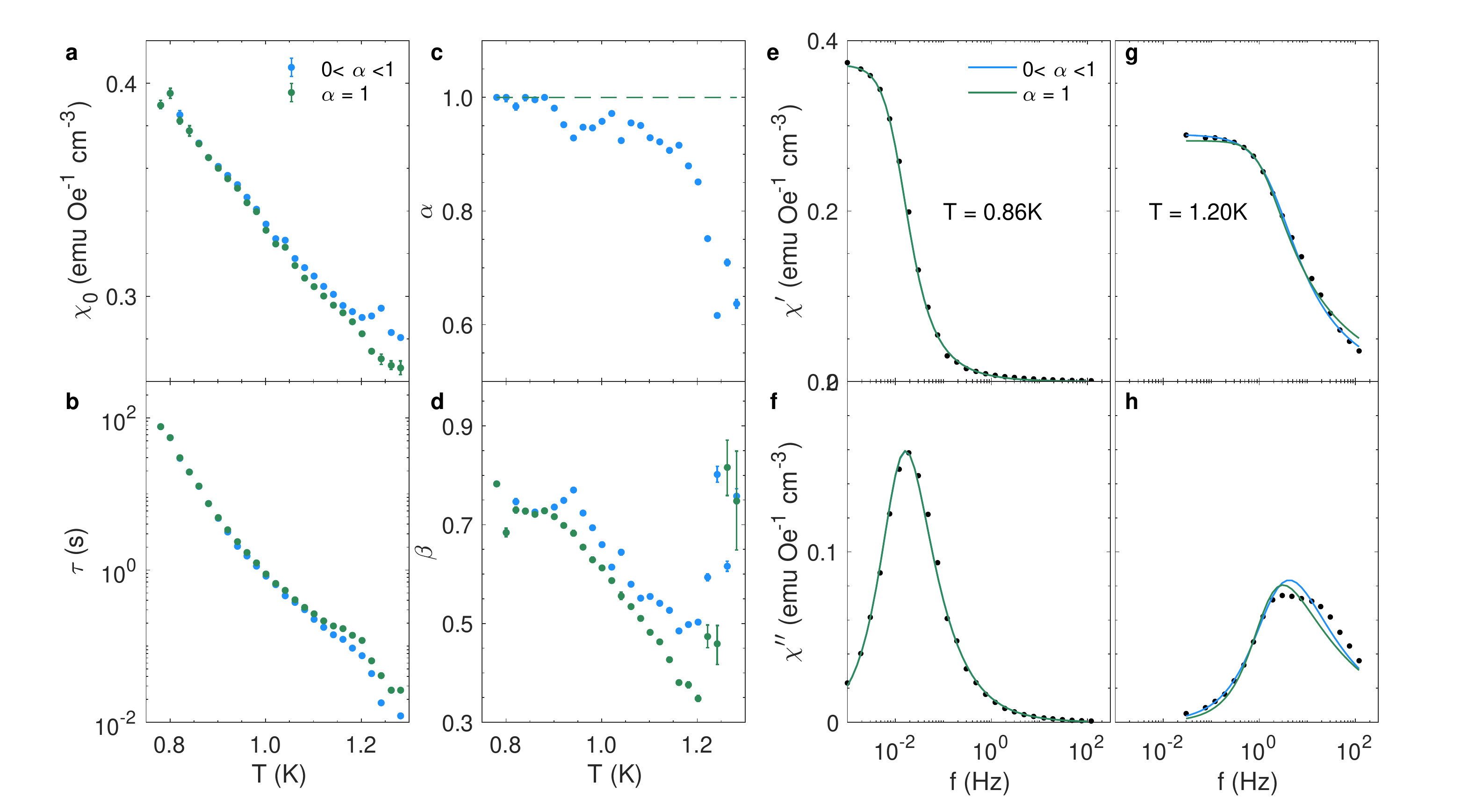}
    \caption{\textbf{Cole-Davidson form as the minimal model to describe the magnetic susceptibility and the breakdown of the single-mode description.} We fit $\chi(f)$ measured by SQUID (main Fig. 2b) using the Havriliak-Negami form of $\chi(\omega=2\pi f)=\frac{\chi_0}{(1+(i\omega\tau)^{\alpha})^{\beta}}(0<\alpha,\beta<1)$ \cite{rosa2015relaxation,kassner2015supercooled,eyvazov2018common}, with $0<\alpha\leq1$ as a free parameter (blue symbols and curves) and $\alpha = 1$ fixed (green symbols and curves). As an empirical form to describe response function in the frequency domain, Havriliak-Negami form is reduced to Cole-Davidson form with $\alpha=1$ (Eqn. 2 in the main text) and further reduced to Debye form if $\alpha=\beta=1$ (Methods). \textbf{a}-\textbf{d}, Fitting parameters $\chi_0$, $\tau$, $\alpha$ and $\beta$. At low temperatures, we first notice that best fit of $\alpha$ is essentially 1 even when it is allowed to vary (panel \textbf{c}, blue). Second, fitting results of the other three parameters are comparable no matter $\alpha$ is fixed or free. On the other hand, $\beta$ evidently deviates from 1. Therefore, the Cole-Davidson form with $\alpha=1$ and $\beta<1$ is the minimal and proper model to describe $\chi(f)$ at low temperatures. \textbf{e}-\textbf{h}, $\chi(f)$ curves at $T=0.86$~K and $T=1.20$~K fit by the two models described in panels (\textbf{a}-\textbf{d}). While both models work equally well at low temperature, varying parameters $\alpha$ and $\beta$ fails to capture the lineshape deviating from a single characteristic frequency at high temperatures, necessitating the analysis with two modes.}
    \label{supfig:NotSingle}
\end{figure}

\newpage
\begin{figure}
    \centering
    \includegraphics[width=17cm]{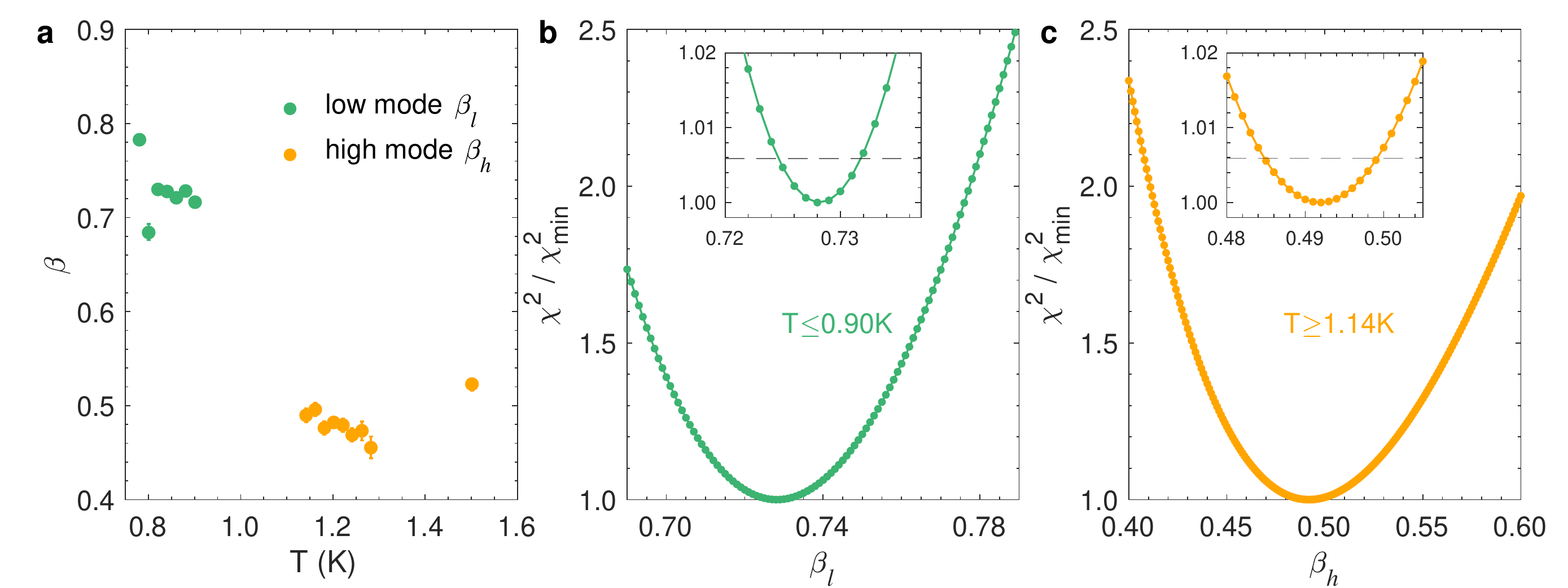}
    \caption{\textbf{Global determination of $\beta_l$ and $\beta_h$.} \textbf{a}, The exponent $\beta$ for the low (green) and high modes (orange) at $T\leq0.9$~K and $T\geq1.14$~K, when the two peaks can be unambiguously separated. Note that $\beta_l$ is essentially $\beta$ in the main Fig.~3c. Both $\beta_l$ and $\beta_h$ manifest no discernible temperature dependence within fluctuations. \textbf{b},\textbf{c}, Normalized $\chi^2$ for global fitting of $\beta_l$ and $\beta_h$. Forcing $\beta_l$ and $\beta_h$ to be temperature independent, we calculated the global $\chi^2$ for best fitting at a given $\beta$ value and found the global optima $\beta_l=0.728\pm0.004$ and $\beta_h=0.492\pm0.008$. (insets) Zoomed views around the optimum with the fitting 1$\sigma$ threshold marked by the dashlines.}
    \label{supfig:Beta}
\end{figure}

\newpage
\begin{figure}
    \centering
    \includegraphics[width=15cm]{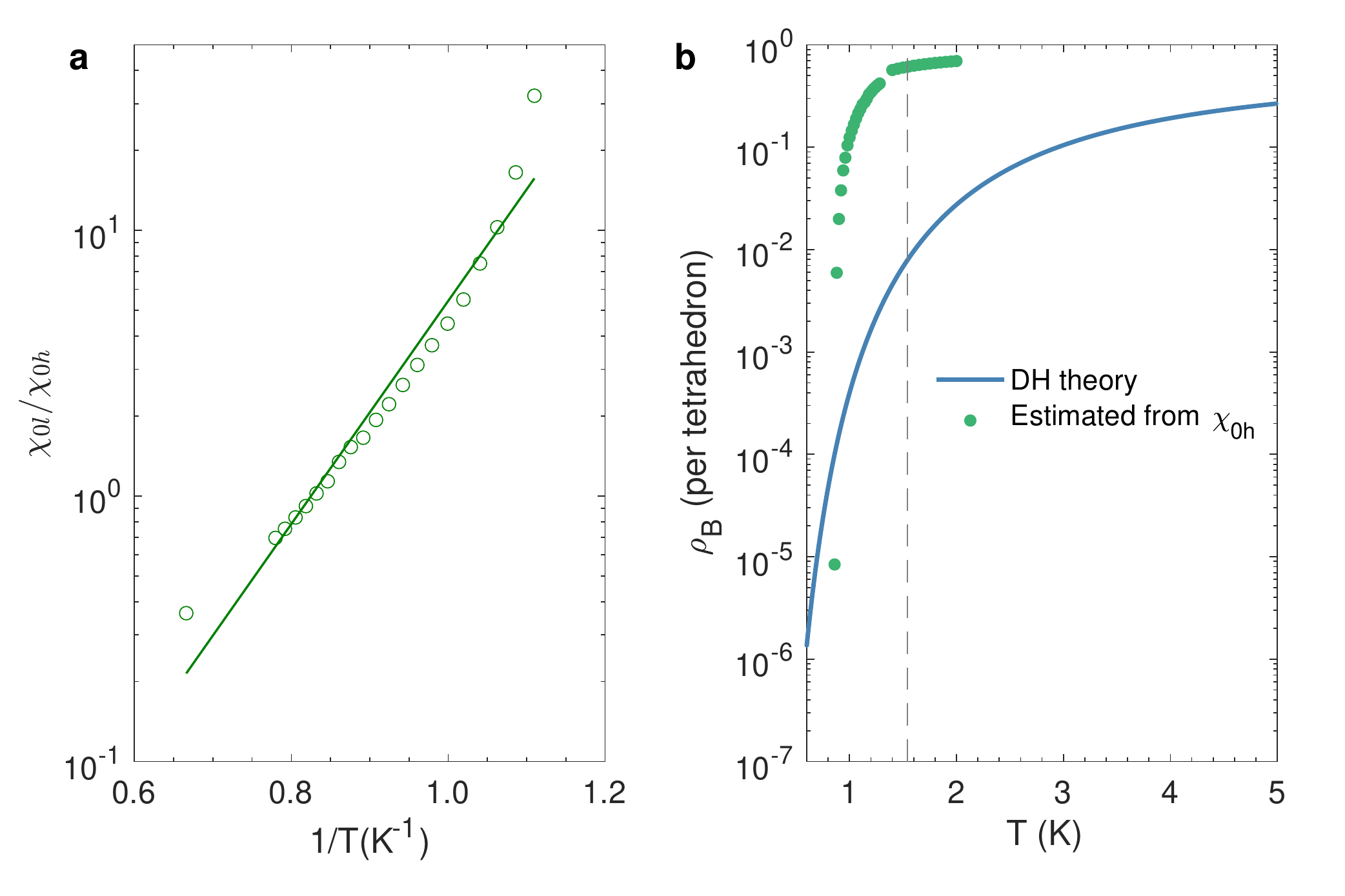}
    \caption{\textbf{Associating the high$\mathbf{-T}$ mode with magnetizable dipoles.} \textbf{a}, In analogy to a thermally populated two-level system, the branching ratio between $\chi_{0l}$ and $\chi_{0h}$ (main Fig.~3a) can be described by $\chi_{0l}/\chi_{0h}=C\exp(\Delta E/T)$, and was plotted against $1/T$. The open circles represent experimental data while the solid curve represent the best fit with $C = 3.4(6)\times10^{-4}$ and $\Delta E=9.7(9)$~K. \textbf{b}, The Bjeruum pair density has been calculated from Debye-H\"uckel theory (blue line), and estimated from $\chi_{0h}$ through $\rho_{\rm B}(T)=T\chi_{0h}(T)/0.49$ (green symbols), derived from Eqn.~S\ref{Eqn: bound pair susceptibility_numerical}. The grey dash line at $T=1.54$~K indicates when the Bjerrum characteristic length $l_{\rm B}$ is estimated to equal the diamond lattice constant $a_{\rm d}$.  Details of Debye-H\"uckel calculation can be found in Supplementary Note 2.}
    \label{supfig:Beta}
\end{figure}

\newpage
\begin{figure}
    \centering
    \includegraphics[width=12cm]{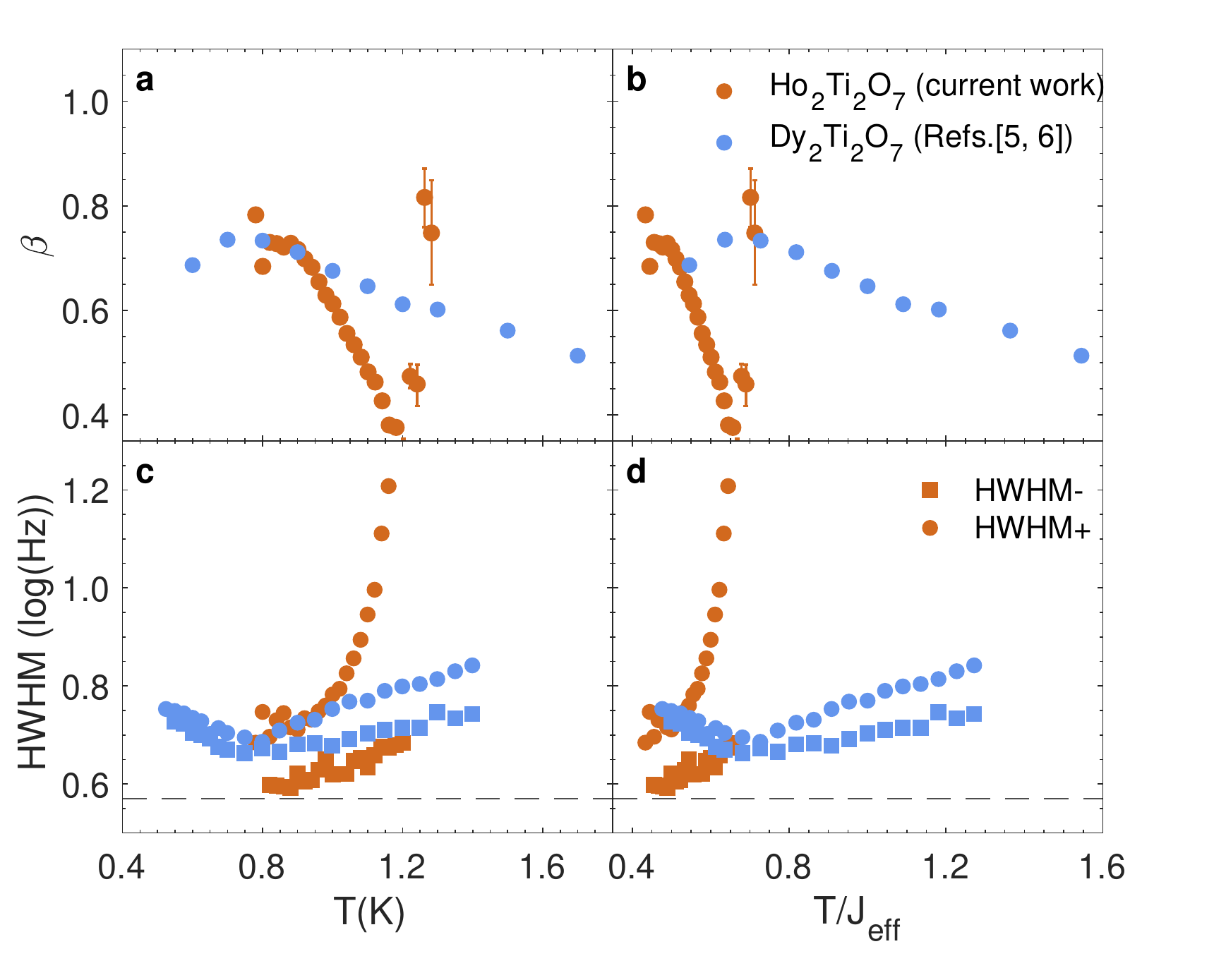}
    \caption{\textbf{Comparative analysis of $\mathbf{\chi(f)}$ lineshape for $\mathbf{Ho_2Ti_2O_7}$ and $\mathbf{Dy_2Ti_2O_7}$.}  As discussed in the main texts and Methods, deviation from a single mode description can be signified by $\beta$ and the asymmetric half-width-half-magnitude (HWHM). We plot $\beta$ (panels \textbf{a},\textbf{b}) and HWHM (panels \textbf{c},\textbf{d}) against temperature (panels \textbf{a},\textbf{c}) and the dimensionless $T/J_{\text{eff}}$ (panels \textbf{b},\textbf{d}) separately for both $\rm Ho_2Ti_2O_7$ (red) and $\rm Dy_2Ti_2O_7$ (blue), with $J_{\text{eff}}=1.8$~K and $1.1$~K, respectively~\cite{melko2004monte}. While data for $\rm Ho_2Ti_2O_7$ are from current work and are identical with what is presented in the main Fig. 3c,d, data of $\rm Dy_2Ti_2O_7$ are from Ref.~\cite{matsuhira2011spin} (panels \textbf{a},\textbf{b}) and Ref.~\cite{yaraskavitch2012spin} (panels \textbf{c},\textbf{d}). \textbf{a},\textbf{b}, In Ref.~\cite{matsuhira2011spin}, two relaxation time scales have been tentatively claimed in $\rm Dy_2Ti_2O_7$ when the Debye form fails to describe the $\chi(f)$ curves. However, we found that the curves presented in Fig. 1 of Ref.~\cite{matsuhira2011spin} can be described by a single mode using the Cole-Davidson form (Eqn. 2 in the main text), while the low temperature $\beta\sim0.7-0.8$ is comparable with the low temperature mode in our measurements of $\rm Ho_2Ti_2O_7$. \textbf{c},\textbf{d}, HWHM- (square) and HWHM+ (circle) for $\rm Ho_2Ti_2O_7$ (red) and $\rm Dy_2Ti_2O_7$ (blue), while the latter is a direct copy of Fig. 2 (inset) in Ref.~\cite{yaraskavitch2012spin}. As far as the scaled $T/J_{\text{eff}}$ is concerned, experimental investigation of $\rm Dy_2Ti_2O_7$ has covered the relevant temperature range corresponding to where the second mode emerges in $\rm Ho_2Ti_2O_7$, while neither $\beta$ nor the asymmetry of lineshape in $\rm Dy_2Ti_2O_7$ exhibits signatures of a distinct relaxation mode as evidently as in $\rm Ho_2Ti_2O_7$.}
    \label{supfig:NotSingle}
\end{figure}

\clearpage

\bibliography{supplementary}